\documentclass[%
preprint,
 amsmath,amssymb,
 sd,
titlepage
]{revtex4-1}
\usepackage{multirow}
\usepackage{graphicx}% Include figure files
\usepackage{dcolumn}% Align table columns on decimal point
\usepackage{bm}% bold math
\usepackage{enumitem}
\usepackage{bigints}
\usepackage[autostyle]{csquotes}
\usepackage[table,xcdraw]{xcolor}
\definecolor{codegreen}{rgb}{0,0.6,0}
\definecolor{codegray}{rgb}{0.5,0.5,0.5}
\definecolor{codepurple}{rgb}{0.58,0,0.82}
\definecolor{backcolour}{rgb}{0.95,0.95,0.92}
\usepackage[T1]{fontenc}
\usepackage[english]{babel}
\MakeOuterQuote{"}
\usepackage{siunitx}
\usepackage[version=4]{mhchem}
\usepackage[acronym]{glossaries}

\makeatletter
\newcommand\reaction@[1]{\begin{equation}\ce{#1}\end{equation}}
\newcommand\reaction@nonumber[1]%
{\begin{equation*}\ce{#1}\end{equation*}}
\newcommand\reaction{\@ifstar{\reaction@nonumber}{\reaction@}}
\makeatother

\newacronym{bp}{BP}{benzophenone}
\newacronym{coin}{CoIn}{conical intersection}
\newacronym{uv}{UV}{ultraviolet}
\newacronym{fwhm}{FWHM}{full width at half maximum }
\newacronym{mbp}{m-BP}{\emph{meta}-methyl benzophenone}
\newacronym{pes}{PES}{potential energy surface}
\newacronym{fc}{FC}{Franck-Condon}
\newacronym{dof}{d.o.f.}{degrees of freedom}
\newacronym{si}{SI}{supporting information}
\newacronym{tdm}{TDM}{transition dipole moment}
\newacronym{qd}{QD}{quantum dynamics}
\newacronym{casscf}{CASSCF}{complete active space self-consistent field}
\newacronym{ci}{CI}{configuration interaction}
\newacronym{caspt2}{CASPT2}{complete active space 2nd order perturbation theory}
\newacronym{xmscaspt2}{XMS-CASPT2}{extended multi-state complete active space 2nd order perturbation theory}
\newacronym{dftmrci}{DFT/MRCI}{density functional theory / multi-reference configuration interaction}
\newacronym{tda}{TDA}{Tamm-Dancoff approximation}
\newacronym{lrdft}{LR-DFT}{linear response density functional theory}
\begin{document}
\title{Simulating Nonadiabatic Dynamics in Benzophenone: Tracing Internal Conversion Through Photoelectron Spectra}

\author{Lorenzo Restaino}
\affiliation{Department of Physics, Stockholm University, Albanova University Centre, SE-106 91 Stockholm, Sweden}
\author{Thomas Schnappinger}%
\affiliation{Department of Physics, Stockholm University, Albanova University Centre, SE-106 91 Stockholm, Sweden}
\author{Markus Kowalewski}%
\email{e-mail: markus.kowalewski@fysik.su.se}
\affiliation{Department of Physics, Stockholm University, Albanova University Centre, SE-106 91 Stockholm, Sweden}

\begin{abstract}
Benzophenone serves as a prototype chromophore for studying the photochemistry of aromatic ketones, with applications ranging from biochemistry to organic light-emitting diodes.
In particular, its intersystem crossing from the first singlet excited state to triplet states has been extensively studied, but experimental or theoretical studies on the preceding internal conversion within the singlet manifold are very rare.
This relaxation mechanism is particularly important because direct population transfer of the first singlet excited state from the ground state is inefficient due to its low oscillator strength.
In this work, we aim to fill this gap by employing mixed quantum classical and full quantum dynamics simulations and time-resolved photoelectron spectroscopy for gas-phase benzophenone and \textit{meta}-methyl benzophenone.
Our results show that nonadiabatic relaxation via conical intersections leads to a linear increase in the population of the first singlet excited state.
This population transfer due to conical intersections can be directly detected by a bifurcation of the photoelectron signal.
In addition, we are able to clarify the role of the third singlet excited state degenerate to the second excited state --- a topic that remains largely unexplored in the existing literature on benzophenone.
\end{abstract}

\maketitle

\section{Introduction}
\Gls{bp} is an organic chromophore frequently employed as a prototype for investigating the photochemistry of more complex aromatic ketones.
This type of chromophore has a rich photochemistry under \gls{uv} light, which involves several nonadiabatic steps: a \gls{coin} between the first two singlet excited states, $S_2$ and $S_1$, and subsequent efficient intersystem crossing processes between the $S_1$ and several triplet states.
Especially due to the latter \gls{bp} and many of its derivatives have significant applications in technological and biological contexts, including organic light-emitting diodes~\cite{Lee2014-iq,Jhulki2016-hi}, \gls{uv} filters~\cite{Li2016-km, Wong2021-gd}, and photosensitizers~\cite{Cuquerella2012-be,Dumont2015-mp}.
The existing literature on the photochemistry of \gls{bp} is extensive, but most experimental and theoretical studies~\cite{Aloise2008-pl,Spighi2014-hv,Sergentu2014-da,Favero2016-ht,Marazzi2016-lc,Shizu2021-ij} focus on the mechanisms leading to direct and indirect population of the lowest triplet excited state, $T_1$, starting from $S_1$~\cite{Marazzi2016-lc}.
However, the $S_1$ state has a negligible cross-section and cannot be efficiently populated directly from the ground state.
The experimental more realistic pathway is to excite the system to higher lying singlet states of $\pi \pi ^*$ character, which have a stronger oscillator strength, and populate $S_1$ via internal conversion.
\begin{figure}[b]
    \centering
    \includegraphics[width=0.7\linewidth]{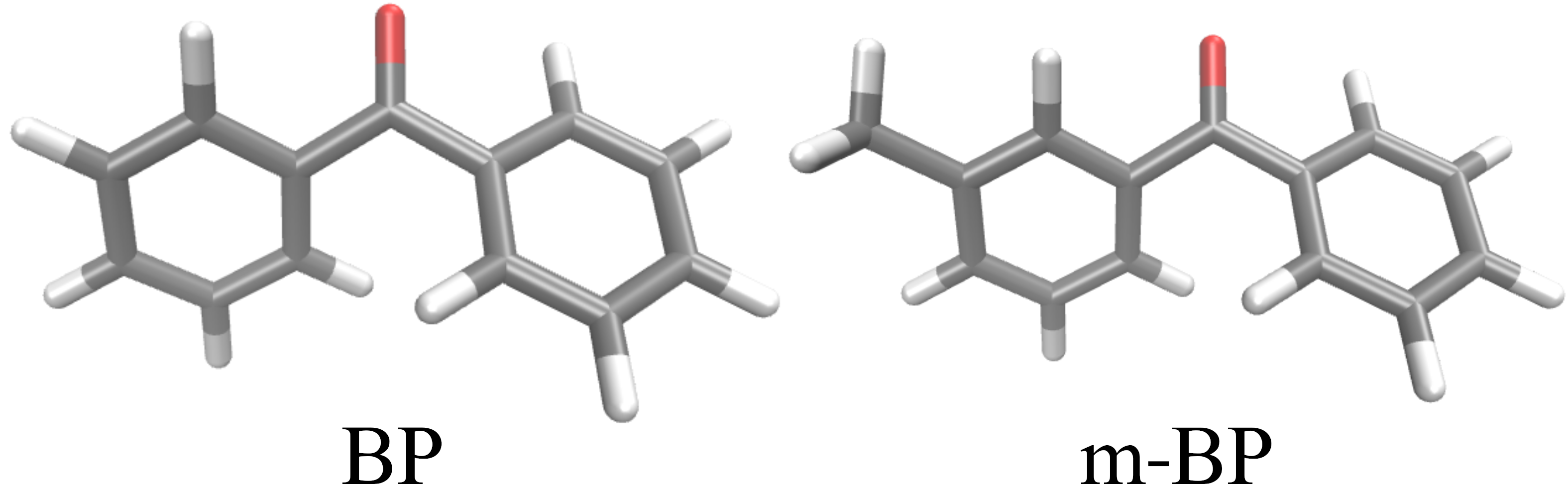}
    \caption{Molecular structures of \acrlong{bp} and \acrlong{mbp}.}
    \label{BP_mBP_mol_struct}
\end{figure}

The $S_2$/$S_1$ \gls{coin} has been the subject of two experimental investigations by Shah et al.~\cite{Shah2004-zh} and Spighi et al.~\cite{Spighi2014-hv}.
In their study, Shah et al. employed time-resolved absorption spectroscopy with a \gls{fwhm} $\approx150$~fs pump pulse centered at 267~nm ($\approx4.66$~eV).
They measured a lifetime of 0.53~ps for the $S_2$ state in acetonitrile following the growth in absorption of the $S_1$ state, and a kinetic rate of $1.7-1.9 \cdot 10^{12}$~s$^{-1}$ for the $S_2 \rightarrow S_1 $ internal conversion.
Spighi et al. investigated the excited state dynamics of gas-phase \gls{bp} via femtosecond and nanosecond resonance-enhanced multiphoton ionization, employing a pump pulse centered at 266~nm ($\approx4.66$~eV), with a pump-probe cross-correlation of approximately 100~fs.
The authors observed an exponential rise and decay of photoelectron bands with a time constant of 150~fs, which they attributed to internal conversion to $S_1$. However, as the authors stated, the multiphoton ionization probe made it impossible to unambiguously assign each band in the photoelectron spectra to a single electronic state.
The resonance-enhanced multiphoton ionization scheme involves several intermediate states providing multiple pathways and resonance conditions,
further complicating the interpretation.

\begin{figure}
    \centering
\includegraphics[width=0.4\linewidth]{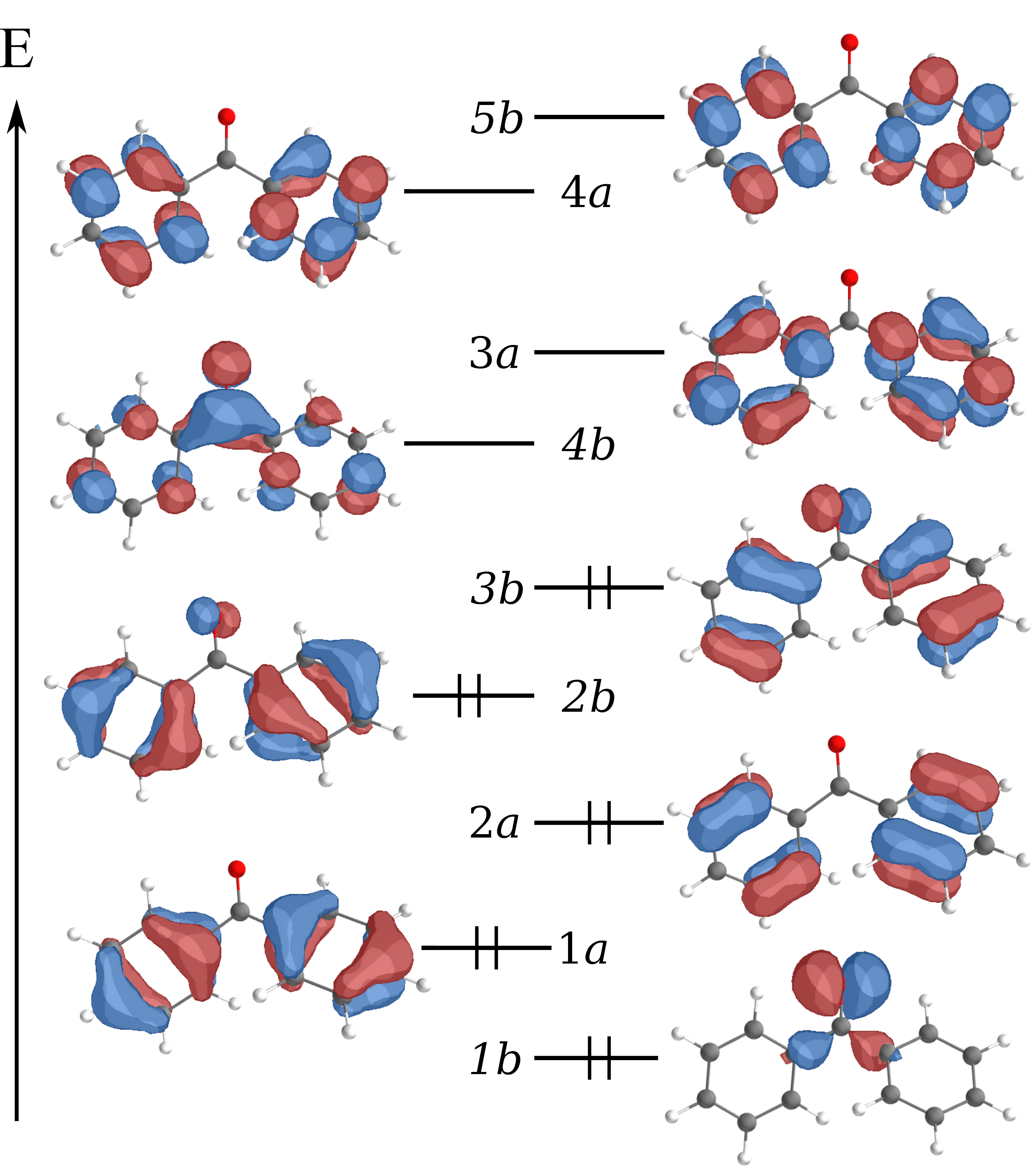}
    \caption{Valence molecular orbital diagram of ground-state \gls{bp} at the CAS(16,15)/ANO-L-VDPZ level of theory. The molecular orbitals are labeled according to $C_2$ symmetry. The numbering for the oxygen lone pair in the diagram is arbitrarily set to 1. An isovalue of 0.05 was used to generate the molecular orbital plots.}
    \label{BP_MO_diagram}
\end{figure}
Both experiments provide evidence of the ultrafast nonradiative decay from the $S_2$ state, indicating the presence of a conical intersection between $S_2$ and $S_1$. However, the temporal resolution of the two experiments cannot resolve the intricate ultrafast excited-state dynamics occurring within the first 100~fs. Moreover, the existence of the $S_3$ state, which is almost degenerate to the $S_2$ state, has not been taken into account in the interpretation of the experimental results.
The $S_1$ state has $n\pi^*$ character and belongs to the irreducible representation $A$ under $C_2$ symmetry. The $S_0 \rightarrow S_1$ transition predominantly involves the promotion of an electron from the oxygen lone pair, $1b$, to the lowest unoccupied molecular orbital (LUMO), $4b$ (see fig.~\ref{BP_MO_diagram}).
Both $S_2$ and $S_3$ have $\pi \pi^* $ character and belong to the irreducible representations $A$ and $B$, respectively. The $S_0 \rightarrow S_2$ transition primarily involves $\pi$ electron excitations from $2b$ to $3b$ and from $2a$ to $3a$; while the $S_0 \rightarrow S_3$ transition involves $\pi$ excitations from $1a$ to $4b$ and from $2b$ to $4a$.
Given that the energy difference between $S_2$ and $S_3$ is predicted on the order of $10^{-2}$~eV, the experimental pump pulse resolution cannot differentiate them.
Consequently, both states can be expected to be populated under experimental conditions~\cite{Shah2004-zh,Spighi2014-hv}, adding another piece to the puzzle.
In this work we determine time-resolved photoelectron spectra based on the excited state dynamics in the singlet manifold of the gas phase \gls{bp} and its methylated derivative \gls{mbp}, displayed in fig.~\ref{BP_mBP_mol_struct}, to show that
(i) the rise in the population of the $S_1$ state is linear, and it is accompanied by a similar linear increase in the corresponding photoelectron band. Beyond that, the internal conversion can be associated with a bifurcation of the photoelectron signal;
(ii) the inclusion of the $S_3$ state reveals ultrafast population transfer to $S_2$, due to the presence of a conical intersection in the vicinity of the \gls{fc} point.

\section{Computational Details \label{sec:Comput_det}}
All calculations were performed in a reproducible environment using the Nix package manager together with NixOS-QChem~\cite{Kowalewski2022-zt} (commit 9b40085ca).

\subsection{\textit{Ab initio} level of theory}
All calculations were performed in gas phase, in $C_1$ symmetry.
Geometry optimizations and frequency analysis were carried out at the $\omega$B97X-D2~\cite{Chai2008-rw,Grimme2006-pj}/6-311G(d)~\cite{Krishnan1980-fy} level of theory using the Gaussian 16 RevC.01~\cite{Frisch2016-rb} quantum chemistry program.
The excited states of \gls{bp} and \gls{mbp} were determined by \gls{lrdft}, \gls{casscf} as well as \gls{dftmrci} theories.
The \gls{lrdft} calculations were carried out within the  \gls{tda}~\cite{Hirata1999-ub} using $\omega$B97X-D4~\cite{Najibi2020-fe}/6-311G(d) level of theory in the ORCA 5.0.4 program package~\cite{Neese2012-bk,Neese2022-ra}.
The state-average SA-\gls{casscf} calculations were performed using OpenMolcas v.24.02~\cite{F_Aquilante2020-ds,Li-Manni2023-ro}, with the ANO-L-VDZP~\cite{Widmark1990-xz} basis set, averaging over 4 states. After the \gls{casscf} calculation, dynamic electron correlation was added using \gls{xmscaspt2}~\cite{Shiozaki2011-zl,Granovsky2011-ar}.

The \gls{ci} expansion for \gls{casscf} scales nearly exponentially with the number of active orbitals and electrons. The full $\pi$-system of BP requires an active space of 16 electrons in 15 orbitals. The full CAS(16,15) active space is shown in the supporting information fig.~S1. This computation is quite time-consuming, and \gls{caspt2} becomes
impractical for the purpose of calculating excited state dynamics.
Although this active space can still be used to benchmark some critical points, it is not suitable for constructing global potential energy surfaces.
For this reason, we employed a \gls{dftmrci} approach to achieve balance between computational costs and accuracy.

Two-dimensional potential energy surfaces, diabatic couplings and \glspl{tdm} of the ground state and three singlet excited states of \gls{mbp} were calculated at the \gls{dftmrci}(2)~\cite{Grimme1999-qv,Neville2022-ht,Costain2024-fo} level using the QTP17~\cite{Jin2018-hm} xc-functional, the QE8 hamiltonian~\cite{Costain2024-fo} and the 6-311G(d) basis set. The quasi-diabatic states and potentials were obtained by a procedure using quasi-degenerate perturbation theory~\cite{Neville2024-or} as implemented in GRACI~\cite{Neville2021-fp}. The active space used in the \gls{dftmrci} calculations is shown in figs.~S3 and S4 of the \gls{si}.
The raw values (energies, couplings, and transition dipole moments) were then interpolated with polyharmonic splines~\cite{Kowalewski2016-ot} onto a fine grid ($N_1=N_2=128$ points), for the wave packet dynamics.

\subsection{Surface Hopping Dynamics}

The molecular dynamics of \gls{bp} and \gls{mbp} were simulated using the SHARC \textit{ab initio} dynamics package version 3.0~\cite{Mai2018-bl,Mai2023-yu}.
The necessary energies, gradients and non-adiabatic couplings were calculated on-the-fly at the \mbox{\gls{tda}/$\omega$B97X-D4/6-311G(d)} level of theory.
In total, 1000 initial conditions for the dynamics simulations were generated based on a Wigner distribution computed from harmonic vibrational frequencies in the optimized ground state equilibrium geometry.
The underlying frequency calculations were performed at \mbox{$\omega$B97X-D2/6-311G(d)} level of theory.
Altogether, 160 initial geometries and velocities, 68 in state $S_2$ and 92 in state $S_3$ were chosen stochastically.

The SHARC \textit{ab initio} surface-hopping algorithm uses a fully diagonal electronic basis~\cite{Mai2013,Mai2015}. The integration of the nuclear motion was done with the Velocity-Verlet algorithm with a maximal time of 0.5~ps using a time step of 0.5~fs.
In each time step, gradients were computed for all states which are closer than 0.001~eV to the active state.
The coefficients of the electronic wave function are propagated on interpolated intermediates with a time step of 0.02~fs applying a local diabatization technique~\cite{Granucci2007-ou} in combination with the WFoverlap code~\cite{Plasser2016} to compute the wave function overlaps.
Decoherence correction was taken into account using the energy-based method of Granucci and Persico with the parameter $\alpha$~=~0.1~a.u.~\cite{Granucci2010-cq}
We restricted our analysis to trajectories with a maximum change in the total energy of 0.3~eV and a maximum change in the total energy per step of 0.2~eV. All trajectories are included in the analysis as long as both criteria are fulfilled. At 500 fs, 55{\%} of the trajectories for \gls{bp} and 50{\%} of the trajectories for \gls{mbp} met the energy conservation criteria.

For each geometry, Dyson norms were calculated between 4 neutral and 4 cationic states through the WFoverlap code~\cite{Plasser2016,Ruckenbauer2016-ep} at the \mbox{\gls{tda}/$\omega$B97X-D4/6-311G(d)} level of theory. Excitation energies and corresponding Dyson norms were employed to generate a photoelectron spectrum, which was then broadened by a Gaussian function with a full width at half maximum of 0.4~eV.

\subsection{Quantum Dynamics Simulations}
The wave packet dynamics of the photoexcited \gls{mbp} was calculated by solving the time-dependent (non-relativistic) Schr\"odinger equation numerically with the Fourier method~\cite{Feit1982-mr,Kosloff1983-yl,Tannor2007-rl}, using the in-house software package QDng~\cite{Kowalewski2024-va}.
To propagate the wave packet on the potential energy surfaces,
we used a Hamiltonian in a diabatic basis and coordinates $q$:
\begin{equation} \label{eq:H}
    \hat{H} = \begin{pmatrix}
            \hat{T} + \hat{V}_{\mathrm{S_0}}(q) & 0 & 0 & 0 \\
            0
            &  \hat{T} + \hat{V}_{\mathrm{S_1}}(q) &  \hat{V}_{\mathrm{S_1}\mathrm{S_2}}(q) & 0 \\
           0 & \hat{V}_{\mathrm{S_1}\mathrm{S_2}}(q) & \hat{T} + \hat{V}_{\mathrm{S_2}}(q) & \hat{V}_{\mathrm{S_2}\mathrm{S_3}}(q) \\
0 & 0 & \hat{V}_{\mathrm{S_2}\mathrm{S_3}}(q) & \hat{T} + \hat{V}_{\mathrm{S_3}}(q)
             \end{pmatrix} \, ,
\end{equation}
where $\hat{T}$ is the kinetic energy operator, $\hat{V}_{\mathrm{S_0}}$, $\hat{V}_{\mathrm{S_1}}$,
$\hat{V}_{\mathrm{S_2}}$ and $\hat{V}_{\mathrm{S_3}}$ are the diabatic potentials for the states $S_0$, $S_1$, $S_2$ and $S_3$,
respectively. The diabatic couplings are given by $\hat{V}_{\mathrm{S_1}\mathrm{S_2}}$ and $\hat{V}_{\mathrm{S_2}\mathrm{S_3}}$.
The kinetic operator is in the G-matrix form~\cite{Stare2003-vi}:
\begin{equation} \label{T_Gmatrix}
    \hat{T} = -\dfrac{\hbar ^2}{2} \sum _{r=1} ^M \sum _{s=1} ^M   \left[ G^{rs} \dfrac{\partial ^2}{\partial q_r \partial q_s} \right] \, ,
\end{equation}
where $M$ is the number of coordinates taken into considerations. The expression in~\eqref{T_Gmatrix} derives from the assumption that the elements of the Wilson G-matrix are constant. The term $G^{rs}$ indicate the element of kinetic energy matrix equal to:
\begin{equation} \label{G_matrix}
    G^{rs} = \sum _{i=1} ^{3N} \dfrac{1}{m_i} \dfrac{\partial q_r}{\partial x_i}\dfrac{\partial q_s}{\partial x_i} \, .
\end{equation}
Here $N$ is the total number of atoms and $m_i$ the mass of atom $i$. In eq.~\eqref{G_matrix}, diagonal entries are reciprocals of generalized reduced masses, and the off-diagonal entry is the kinetic coupling term. The calculated values are given here in atomic units: $G_{rr} = 0.0001447$, $G_{ss}= 0.00016924$ and $G_{rs} =-5.101169 \cdot 10^{-6}$~$\mathrm{au}^{-1}$. The short iterative Lanczos scheme~\cite{Lanczos1950-ex, Park1986-hj, Hochbruck2006-pa} was used
to propagate the nuclear wave packet and the Hamiltonian form eq.~\ref{eq:H}
with a time step of $\Delta t = 1$~au (24.2~as).
The total wave function was expressed as a linear combination
\begin{equation}
    \Psi = \sum _{i=0} ^{4} c_i \phi _i
    \end{equation}
with $c_0 = c_1  = 0$ , $c_2 =\sqrt{0.4}$ and $c_3 = \sqrt{0.6}$. These values were specifically chosen to replicate the initial populations of the surface hopping dynamics.
\section{Results and Discussion}
\subsection{Benchmark of Vertical Excitations}
Before we introduce the molecular dynamics and the simulated time-resolved photoelectron spectra for \gls{bp} and \gls{mbp}, we discuss the vertical UV excitations and ionization energies of the two molecules. Table~\ref{vert_energ_benchmark_BP} provides a benchmark of calculated vertical excitations for \gls{bp} against the experimental values taken from reference~\cite{Itoh1985-jk}. This includes the level of theory used in surface hopping dynamics (\gls{tda}), the level of theory used to construct \glspl{pes} (\gls{dftmrci}), and the multireference method (CASPT2) that employs the full $\pi$ system of \gls{bp} (CAS(16,15) with 4-state averaging). An even more extensive benchmark, including CASPT2 with different sizes of the active space and RASPT2 calculations, can be found in the \gls{si} in tables~S1 to S4. Additional computational details are given in the table caption and in section~\ref{sec:Comput_det}.
\begin{table}
\caption{Comparison between vertical excitation energies and their corresponding oscillator strength of \gls{bp}, calculated at \gls{tda}/$\omega$B97x-D4/6-311G*, \gls{dftmrci}(2)/QTP17/QE8/6-311G* and XMS-CASPT2/CAS(16,15)/ANO-L-VDPZ levels of theory.}
\label{vert_energ_benchmark_BP}
\begin{tabular}{lcccccccc}
\hline
        & \multicolumn{4}{c}{Energies (eV)} &  & \multicolumn{3}{c}{Oscillator Strength} \\ \cline{2-5} \cline{7-9}
        & TDA & DFT/MRCI & CASPT2 & Exp & & TDA & DFT/MRCI & CASPT2  \\ \cline{1-5} \cline{7-9}
$S_{1}$ & 4.04 & 4.04 & 3.65 & 3.61$^a$ & &  0.0010 & 0.0040      & 0.0007     \\
$S_{2}$ & 5.36 & 4.94 & 4.60 & 4.40$^a$ & &  0.015  & 0.0213      & 0.0036   \\
$S_{3}$ & 5.40 & 4.98 & 4.62 & 4.40$^a$ & &  0.015 & 0.0114      & 0.0023       \\ \hline
\multicolumn{8}{l}{$^{a}$ low pressure vapor, ref.~\cite{Itoh1985-jk}}
\end{tabular}
\end{table}
As expected, the CASPT2 calculations predict vertical excitations very close to the experimental values and are therefore well suited as a general criterion for validating the other theoretical methods applied. In contrast, \gls{dftmrci} and especially \gls{tda} calculations generally overestimate the excitation energies.
The \gls{dftmrci} results are slightly more accurate, but still produce vertical transitions that are systematically blueshifted by approximately 0.4~eV. Meanwhile, \gls{tda} calculations tend to overestimate $\pi \pi ^*$ transitions by nearly 1 eV.
Nonetheless, all methods capture the near degeneracy between $S_2$ and $S_3$. Both \gls{dftmrci} and \gls{tda} predict an energy difference between $S_3$ and $S_2$ of 0.04~eV. According to CASPT2 calculations, the energy difference between $S_2$ and $S_1$ at the \gls{fc} point is 0.95~eV, which is slightly above the experimental value of 0.79~eV. Likely due to error compensation, the \gls{dftmrci} method predicts a 0.9~eV gap, performing better than CASPT2.
The \gls{tda} calculations yield the largest energy gap at 1.32~eV. This overestimation of the $S_2$/$S_1$ energy gap may result in a slower population transfer in the surface hopping dynamics. All methods consistently predict the right trend for the oscillator strengths, identifying $S_0 \rightarrow S_1$ as the weakest transition.

\begin{figure}
    \centering
\includegraphics[width=0.5\linewidth]{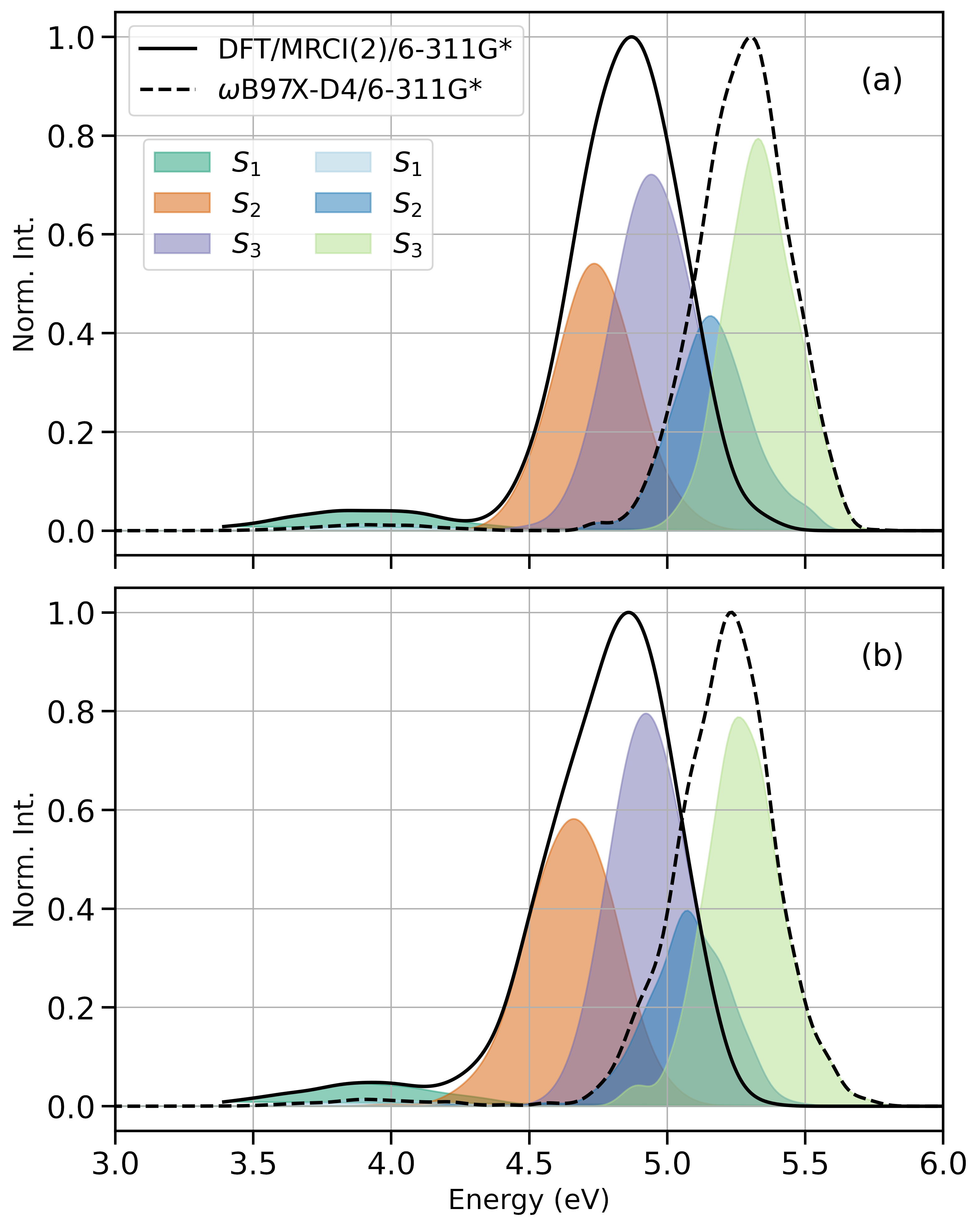}
    \caption{Unshifted UV spectra of benzophenone (a) and \textit{meta}-methyl benzophenone (b) calculated over their respective Wigner distributions with \gls{tda} (dashed line) and \gls{dftmrci} (solid line).}
    \label{wigner_spectrum}
\end{figure}

\begin{table}
\caption{Comparison between vertical excitation energies and their corresponding oscillator strength of \gls{mbp}, calculated at \gls{tda}/$\omega$B97x-D4/6-311G* and \gls{dftmrci}(2)/QTP17/QE8/6-311G* levels of theory.}
\label{vert_energ_benchmark_mBP}
\begin{tabular}{lccccc}
\hline
        & \multicolumn{2}{c}{Energies (eV)} &  & \multicolumn{2}{c}{Oscillator Strength} \\ \cline{2-3} \cline{5-6}
        & TDA & DFT/MRCI & & TDA & DFT/MRCI  \\ \cline{1-3} \cline{5-6}
$S_{1}$ & 4.04 & 4.00 &    & 0.0010  & 0.0023  \\
$S_{2}$ & 5.25 & 4.80 &    & 0.0220   & 0.0279 \\
$S_{3}$ & 5.39 & 4.92 &    & 0.0133  & 0.0124 \\ \hline
\end{tabular}
\end{table}
Table~\ref{vert_energ_benchmark_mBP} shows the vertical excitation energies calculated for \gls{mbp}. Although we were unable to find any experimental UV spectrum data for gas-phase \gls{mbp} to directly compare with our calculations, we anticipate that it will not differ significantly from that of the parent molecule. \gls{tda} and \gls{dftmrci} calculations yield the same value for the $S_0 \rightarrow S_1$ transition as observed for \gls{bp}, suggesting that the methyl group in \textit{meta} has negligible influence of on the CO $\pi$ bond.
The predicted energy gap between $S_3$ and $S_2$ is 0.14~eV with \gls{tda} and 0.12~eV with \gls{dftmrci}, where the splitting arises mainly from a stabilization effect on the $S_2$ state. The energy gap between $S_2$ and $S_1$ at the \gls{fc} is calculated to be smaller for \gls{mbp} than \gls{bp}.
\gls{dftmrci} calculations place it at 0.8~eV, whereas \gls{tda} estimates it at 1.21~eV. Similarly to \gls{bp}, the overestimation of the energy difference between $S_2$ and $S_1$ may slow down the population transfer during the surface hopping dynamics.
The energies calculated over a Wigner distribution (see section~\ref{sec:Comput_det} for more details) have been convoluted with a Gaussian (FWHM = 0.4~eV) to produce the UV spectra of \gls{bp} and \gls{mbp}, as shown in fig.~\ref{wigner_spectrum} (a) and (b), respectively.
Consistent with the findings at the \gls{fc} point presented in tables~\ref{vert_energ_benchmark_BP} and~\ref{vert_energ_benchmark_mBP}, the optically brighter region of the \gls{tda} spectrum shows a blue shift of 0.4~eV compared to \gls{dftmrci}. There are no notable distinctions between \gls{bp} and \gls{mbp}.

\begin{figure}
    \centering
\includegraphics[width=0.5\linewidth]{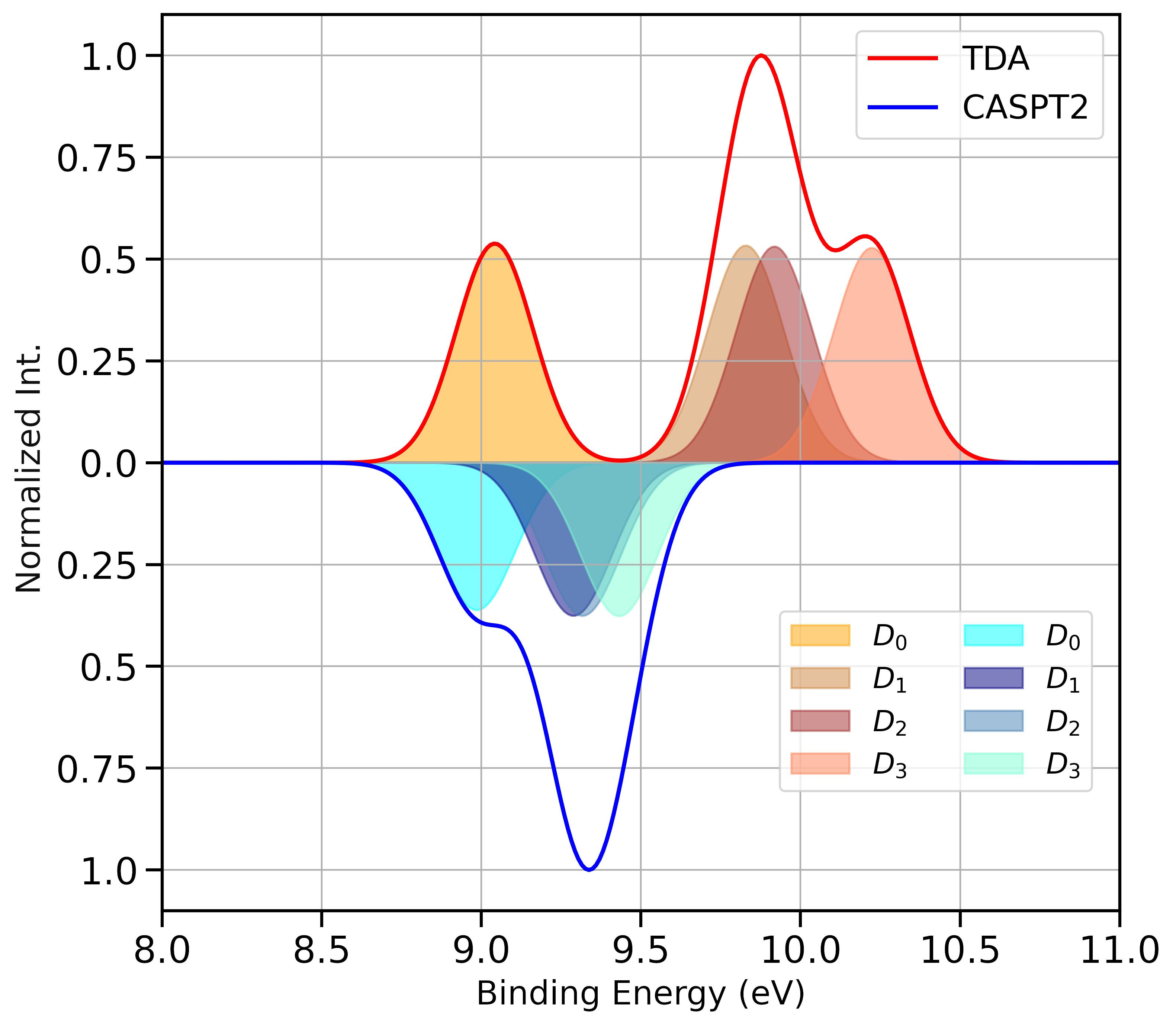}
    \caption{Static photoelectron spectrum of ground-state \gls{bp} calculated with \gls{tda}/$\omega$B97X-D4 (red color scheme) and XMS-CASPT2 (blue color scheme). To prevent overlapping features, the CASPT2 spectrum is shown up-side-down.}
    \label{Static_BP_PS}
\end{figure}

The ionization spectrum of gas-phase \gls{bp} has been the subject of several experimental and theoretical studies~\cite{Centineo1978-nq,McAlduff1979-cn,Khemiri2015-ok,Gouid2019-lu}.
The experimental spectrum measured by Centineo et al.~\cite{Centineo1978-nq} shows an overlapping band in the 8-10~eV range with two distinguishable peaks at approximately 9.0 and 9.45~eV, along with two shoulders at higher energies.
The calculated static photoelectron spectrum of \gls{bp} is shown in fig.~\ref{Static_BP_PS}. The CASPT2 calculations including the complete $\pi$ system of \gls{bp} are in good agreement with the experiment.
The lack of a shoulder at higher binding energies is due to the absence of the corresponding electronic states in the SA-CASSCF calculations, which averaged over four states of the cation. The photoelectron spectrum calculated with \gls{tda} accurately models the $D_0$ state, but it overestimates the ionization energies for $D_1$, $D_2$, and $D_3$, which are blueshifted by approximately 0.5~eV.

Overall, although there are some quantitative differences, we are convinced that \gls{tda} and the chosen functional are adequate to describe the mixed quantum classical dynamics of \gls{bp} and \gls{mbp}.

\subsection{Semiclassical Dynamics and Time-resolved Photoelectron Spectra}
Having validated the \textit{ab initio} method, we will now discuss the excited-state dynamics of \gls{bp} and \gls{mbp}, starting with the parent molecule.
Note that the number of trajectories satisfying the total energy conservation criteria decreases with time. This time-dependent number of trajectories was used to construct the photoelectron spectra and the populations of the singlet manifold. The on-the-fly selection criteria are discussed in section~\ref{sec:Comput_det}.
Upon excitation to the $S_2$ or $S_3$ state, for all analyzed trajectories nonadiabatic processes are observed.
\begin{figure}
    \centering
\includegraphics[width=\linewidth]{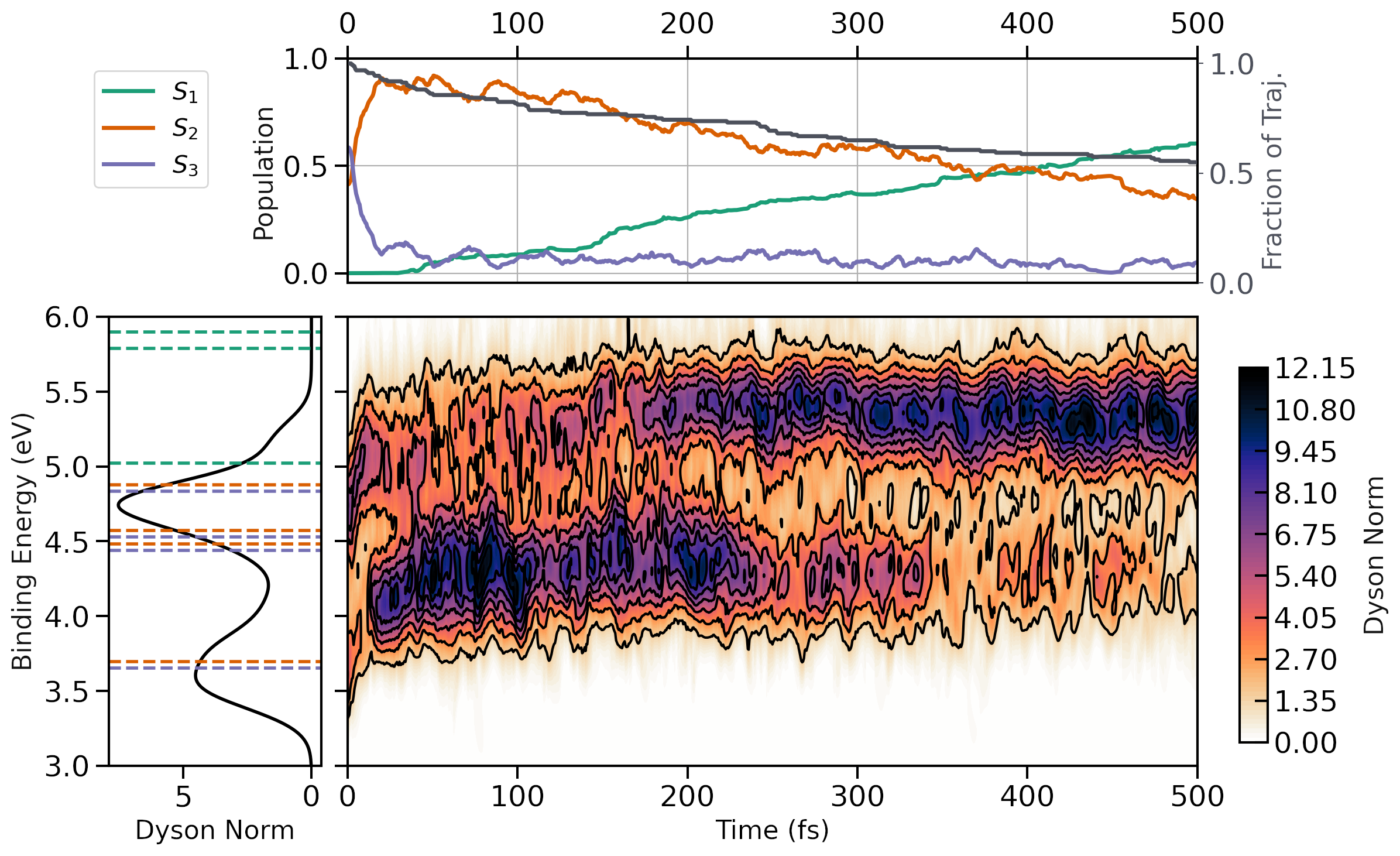}
    \caption{Time-resolved photoelectron spectrum of \gls{bp} computed at the \gls{tda}/$\omega$B97X-D4/6-311G* level of theory from the combined trajectories of $S_2$ and $S_3$. Upper panel: time evolution of the average adiabatic populations along the \gls{bp} trajectories. The second y-axis shows the fraction of trajectories (grey line) meeting the total energy conservation criteria at each time step, used for calculating the spectrum and time-dependent populations of the singlet manifold. A unitary fraction corresponds to 160 trajectories, with a minimum fraction of $\approx 0.55$ equivalent to 88 trajectories. Lower left panel: temporal slice of the time-resolved photoelectron spectrum selected at 0~fs. Dashed lines indicate binding energies at the Frank-Condon point, using the color scheme of the population plot. Lower right panel: full time-resolved photoelectron spectrum.}
    \label{BP_tr_PS_spectrum}
\end{figure}

The time evolution of the average adiabatic populations along the \gls{bp} trajectories is shown in the top panel of fig.~\ref{BP_tr_PS_spectrum}. The second y-axis represents the fraction of trajectories (gray line) that satisfy the total energy conservation criteria at each time step. The trajectories reveal a rapid exponential decay of the $S_3$ population to $S_2$ within the initial 10~fs of the dynamics, and the nonadiabatic relaxation is completed within 50~fs.
The strong depletion in the $S_3$ population suggests that $S_3$ and $S_2$ are strongly coupled and that there is a favorable gradient towards $S_2$.
After 40~fs, an almost linear increase in the $S_1$ population can be observed, along with a corresponding decrease in the $S_2$ population. As more population goes into $S_1$ of $n\pi ^*$ anti-bonding character, an elongation of the CO bond can be observed, as shown in fig.~S6 of the \gls{si}. The normal modes analysis of the SHARC trajectories within the first 150~fs reveals that the dihedral movement between the two aryl rings is predominant. Nevertheless, a direct correlation between the dihedral oscillations and the photoelectron observable could not be established. The normal mode analysis is available in the \gls{si} in fig.~S5.

The lower left panel of fig.~\ref{BP_tr_PS_spectrum} displays a snapshot of the time-resolved photoelectron spectrum taken at time zero. The dashed lines indicate the binding energies at the \gls{fc} point, with the same color scheme as the population plot in fig.~\ref{BP_tr_PS_spectrum}. The snapshot spectrum consists of two bands, approximately centered at 3.6 and 4.7~eV.
With the temporal evolution of the populations in mind, we are now able to discuss the time-resolved photoelectron spectrum of \gls{bp}, shown in the lower right panel of fig.~\ref{BP_tr_PS_spectrum}. Note that due to the different characteristic of the probe (multiphoton vs single photon ionization), we cannot directly compare our results to the experiment of Spighi et al.~\cite{Spighi2014-hv}.
During the initial 15~fs of the dynamics, a distinct blueshift of the two bands is observed in the time-resolved photoelectron spectrum, along with a notable inversion in intensity. The band originating at 3.6~eV experiences a blueshift of approximately 0.4~eV and an increase in intensity, while the band at 4.7~eV experiences a blueshift of approximately 0.3~eV and a decrease in intensity.
The internal conversion from $S_3$ to $S_2$ is responsible for this feature of the time-resolved photoelectron spectrum. The observed intensity exchange matches the exponential decay of the $S_3$ population. Moreover, the blueshift suggests a favorable gradient towards $S_2$. After 50~fs, the bifurcation of the band centered at 3.6~eV becomes increasingly more evident. The signal covers a broad spectral range, from 4 to 5.5~eV. Ultimately, two distinct bands can be identified after 190~fs.
As will be shown later, this spectral feature is even more pronounced in the case of \gls{mbp}. In addition to the splitting of the signal, a distinctive change in intensity is also observed. In particular, the intensity of the new band emerging at 5.3~eV increases linearly, while the other decreases steadily. The nearly linear increase in the intensity profile of the new band is consistent with the time-dependent evolution of the $S_1$ population and its timescale. Therefore, we attribute the bifurcation in the photoelectron signal to the population transfer from $S_2$ to $S_1$.

\begin{figure}
    \centering
\includegraphics[width=\linewidth]{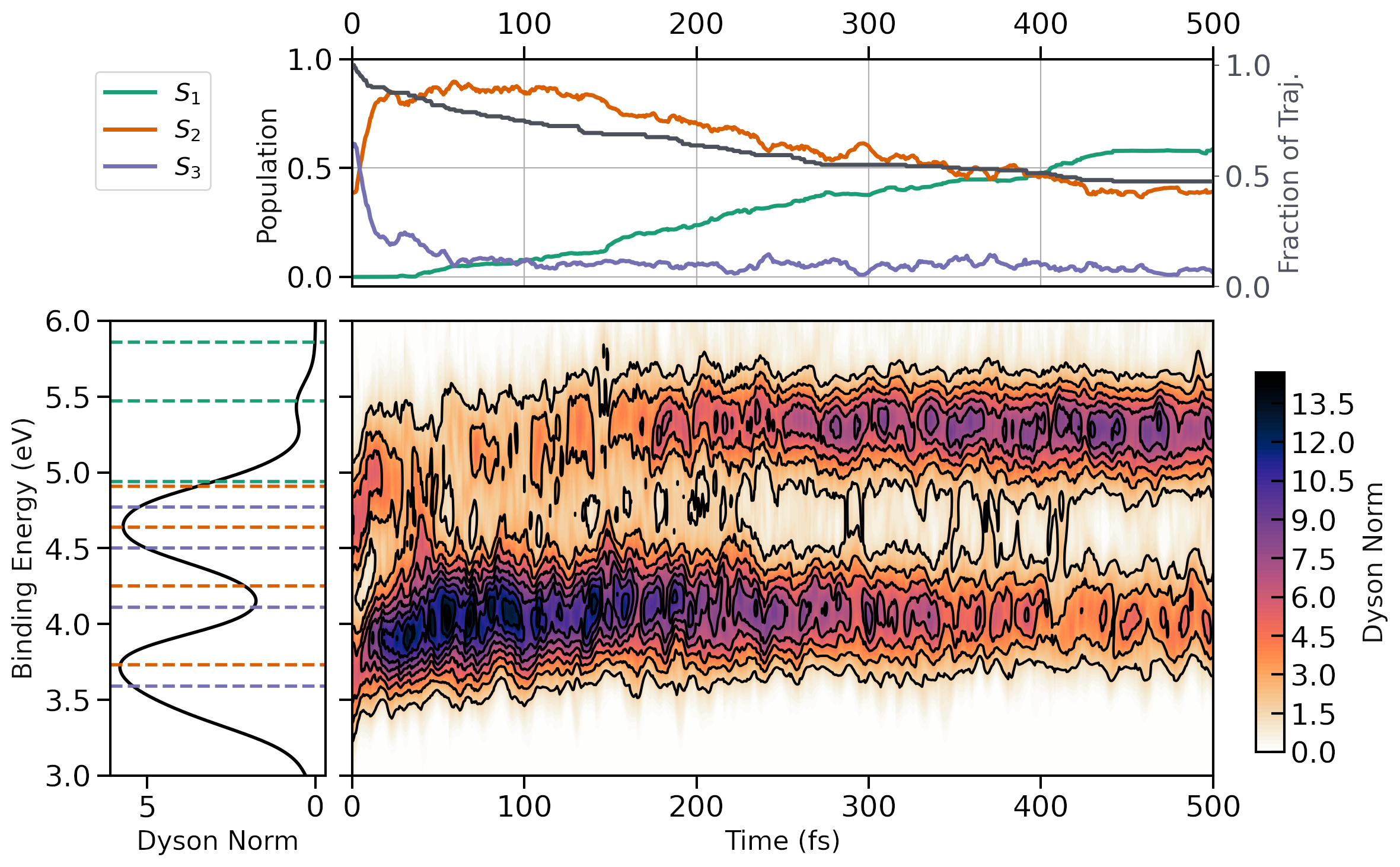}
    \caption{Time-resolved photoelectron spectrum of \gls{mbp} computed at the \gls{tda}/$\omega$B97X-D4/6-311G* level of theory from the combined trajectories of $S_2$ and $S_3$. Upper panel: time evolution of the average adiabatic populations along the \gls{mbp} trajectories. The secondary y-axis shows the fraction of trajectories (grey line) meeting the total energy conservation criteria at each time step. A unitary fraction corresponds to 160 trajectories, with a minimum fraction of $\approx 0.5$ equivalent to 76 trajectories. Lower left panel: temporal slice of the time-resolved photoelectron spectrum selected at 0 fs. Dashed lines indicate binding energies at the \gls{fc} point, maintaining the color scheme of the population plot. Lower right panel: full time-resolved photoelectron spectrum.}
    \label{mBP_tr_PS_spectrum}
\end{figure}

The top panel of fig.~\ref{mBP_tr_PS_spectrum} shows the time evolution of the average adiabatic populations along the \gls{mbp} trajectories. Although the observed features in the population dynamics are very similar to the parent molecule, some small differences also emerge.
The depletion of the $S_3$ population remains exponential. However, a comparison between the two molecules reveals that, during the first 60~fs, the population transferred from $S_3$ to $S_2$ is about 10{\%} greater in \gls{bp} than in \gls{mbp}. After this initial phase, the populations in $S_3$ for both molecules become approximately the same. These features can be better appreciated in fig.~S9 of the \gls{si}. These differences can be attributed to a larger energy gap between $S_3$ and $S_2$, resulting in weaker coupling of the states in \gls{mbp}. After a simulation time of 500~fs, 60{\%} of the total population reside in $S_1$ for both molecules, we conclude that the presence of a methyl group in \textit{meta} position does not significantly affect internal conversion from $S_2$ to $S_1$. Similarly to the parent molecule, there is a direct correlation between the increase in population of $S_1$ and the elongation of the CO bond (fig.~S6 in the \gls{si}).

The spectral snapshot of the time-resolved photoelectron spectrum at 0.0~fs, shown in the left panel of fig.~\ref{mBP_tr_PS_spectrum}, displays an intensity profile different from that of \gls{bp}. In particular, the bands centered at 3.6 and 4.7~eV now show an equal intensity.
The time-resolved photoelectron spectrum of \gls{mbp} is displayed in fig.~\ref{mBP_tr_PS_spectrum}. As in the \gls{bp} situation, the \gls{mbp} spectrum exhibits a blueshift of $\approx 0.4$~eV of these two bands during the initial 15~fs of dynamics, paired with an exponential change in their intensity.
Following this, the signal at 3.9~eV splits, resulting in a linear rise in the intensity of the $S_1$ signal, while the intensity of the $S_2$ signal decreases.
Similarly to the parent molecule, the blueshift, accompanied by the exponential decay/rise in the intensity profile, mirrors the nonadiabatic relaxation from $S_3$ to $S_2$. The signal bifurcation and the linear increase in intensity of the newly emerged band correspond to the nonadiabatic relaxation from $S_2$ to $S_1$, and they appear more pronounced here than \gls{bp}. There is no clear connection between the dihedral oscillations of the aryl rings and the photoelectron spectrum.
\subsection{Construction of Potential Energy Surfaces}
The construction of diabatic \glspl{pes} for \gls{bp} was unattainable due to numerical problems in the diabatization process, attributed to the degeneracy of $S_2$ and $S_3$, and the resulting coupling already at the \gls{fc} point.
To address this issue, we opted for a methylated benzophenone as a candidate to construct diabatic \glspl{pes} and carry out subsequent \gls{qd} simulations.
Indeed, despite the presence of a methyl group, the photochemistry of \gls{mbp} does not change significantly, making the two systems quite comparable, as demonstrated by the \gls{uv} spectrum based on a Wigner sampling in fig.~\ref{wigner_spectrum}.
Positioning the methyl group in \textit{meta} yields a sufficiently large splitting between the $S_3$ and $S_2$ states enough to reduce their coupling at the \gls{fc} point, without introducing a strong steric effect on the relaxation dynamics.
This, in turn, facilitates the construction of diabatic \glspl{pes}. Moreover, asymmetric substitution of the aryl rings breaks the $C_2$ symmetry of \gls{bp}.

The number of vibrational \gls{dof} for \gls{mbp} is $3N-6=75$, where $N$ is the number of atoms in the molecule. In grid-based methods, the size of the nuclear wave function increases exponentially with the number of nuclear d.o.f., which makes the task computationally prohibitive to solve unless we restrict ourselves to the reactive subspace of the multidimensional problem.
In such cases, a choice of reaction coordinate space is required. To describe the initial dynamics associated with the transfer of population from $S_2$ to $S_1$, the difference vector between the \gls{fc} and \gls{coin} geometries was selected as the primary coordinate:
\begin{equation}
    \bm{v}_1 = \bm{R}^{\mathrm{FC}} - \bm{R}^{\mathrm{CoIn}} \, .
\end{equation}
This ensures the presence of the $S_2$/$S_1$ \gls{coin} in the coordinate space by construction. The primary motion of this coordinate qualitatively describes a ring-puckering vibration. The analysis of the \gls{bp} and \gls{mbp} SHARC trajectories revealed that, within the first 150~fs, the dominant motion involved the dihedral angle between the two aryl rings. The normal mode analysis is available in the \gls{si} in fig.~S5. For this reason, the vector describing this motion was chosen as the second coordinate ($\bm{v}_2$). The two vectors appointed as reaction coordinates are depicted in \gls{si} in fig.~S7.

\begin{figure}
    \centering
    \includegraphics[width=\linewidth]{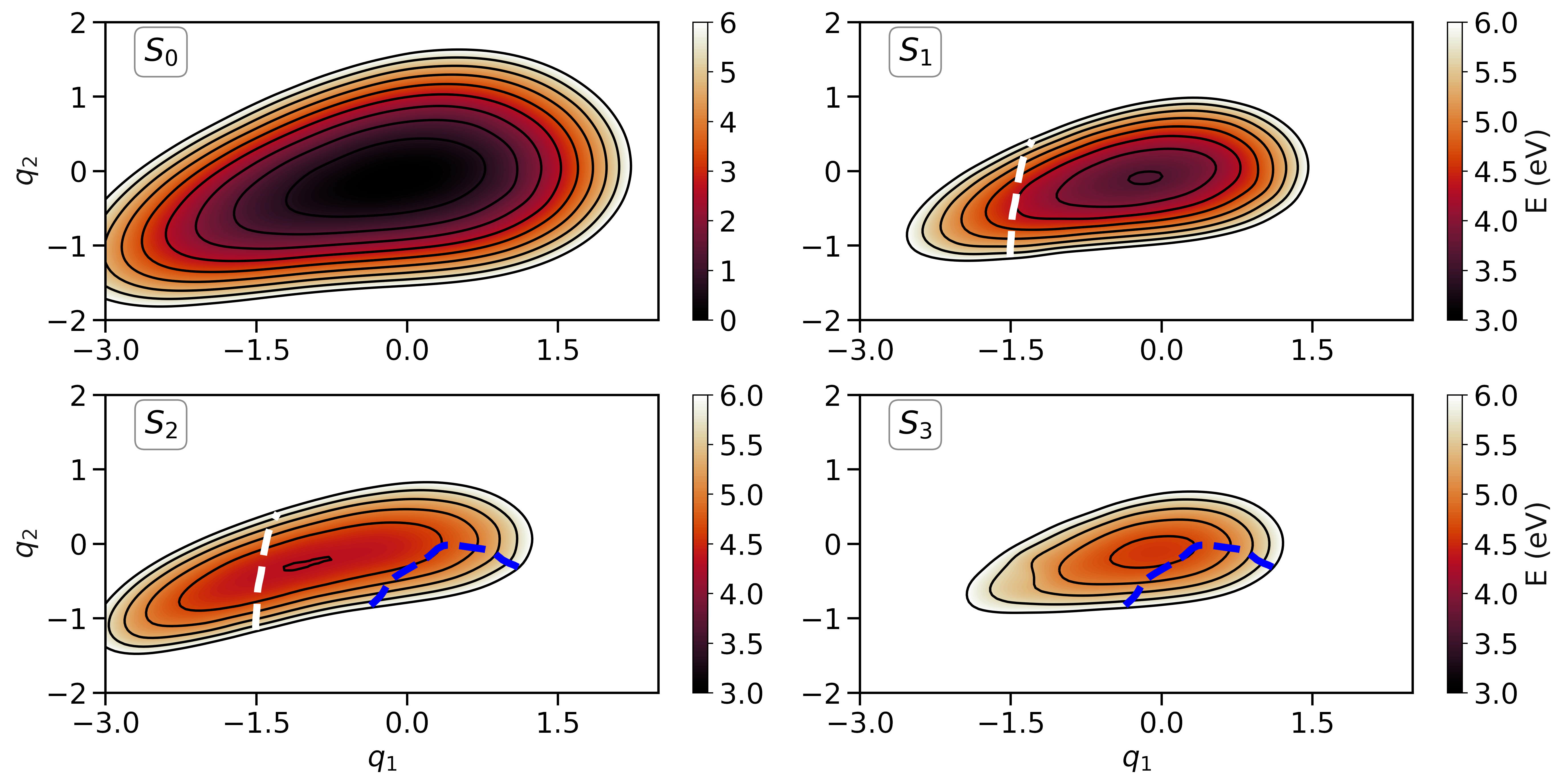}
    \caption{2D diabatic potential energy surfaces of \gls{mbp}.The white dashed line indicates the diabatic intersection seam between $S_2$ and $S_1$ including the optimized \gls{coin}. A blue dashed line indicates the diabatic intersection seam between $S_3$ and $S_2$.}
    \label{meta_BP_dPES}
\end{figure}

After removing the purely translational and rotational components from the two vectors by satisfying the Eckart conditions~\cite{Eckart1935-oy,Louck1976-ji}, we orthonormalized the two coordinates using L{\"o}wdin's orthogonalization scheme.
We then calculated a 2D rectangular grid by scanning the two vectors using the \gls{fc} point as reference geometry:
\begin{equation}
    \bm{R}_{xyz} ^{new} =  \bm{R}_{ref} + q_1 \bm{v}_1 + q_2 \bm{v}_2
\end{equation}
with coefficients $q_1$/$q_2$ in the range [-3, 3]. We conducted an unrelaxed scan of the reaction coordinates. This choice is supported by the timescale of the process we intend to describe, specifically the early sub-150-fs dynamics, during which we assume that additional relaxation does not play a dominant role.
The diabatic \glspl{pes} for \gls{mbp} are displayed in fig.~\ref{meta_BP_dPES}.
\subsection{Quantum Dynamics}

The mixed quantum classical excited-state dynamics based on independent trajectories provides a clear picture of the nonadiabatic relaxation events in \gls{bp} and \gls{mbp}. The surface hopping methodology has the ability to tackle the full dimensionality of the nuclear problem; however, it cannot correctly describe coherent superpositions of electronic states.
In fact, all dynamics start on a single adiabatic electronic state and not a superposition, such as one created by a laser pulse. Moreover, it does not properly account for decoherence in the nuclear subspace, caused by the nuclear motion.
In contrast, an explicit full quantum dynamics simulation correctly describes coherent superpositions and the wave packet branching near a conical intersection, producing a short-lived electronic coherence.
Information about the energy gap between the coupled electronic states is encoded in the oscillation period of these transient electronic coherences, and their detection can be used as a direct signature of \glspl{coin}.
However, the grid-based approach we use is only possible for reduced geometric subspaces and not for the full 3$N$-6 degrees of freedom. For this reason, we have constructed global \glspl{pes} in a reduced subspace of the 3$N$-6 nuclear \gls{dof} of \gls{mbp}, and used them for full \gls{qd} simulations.

\begin{figure}
    \centering
    \includegraphics[width=0.5\linewidth]{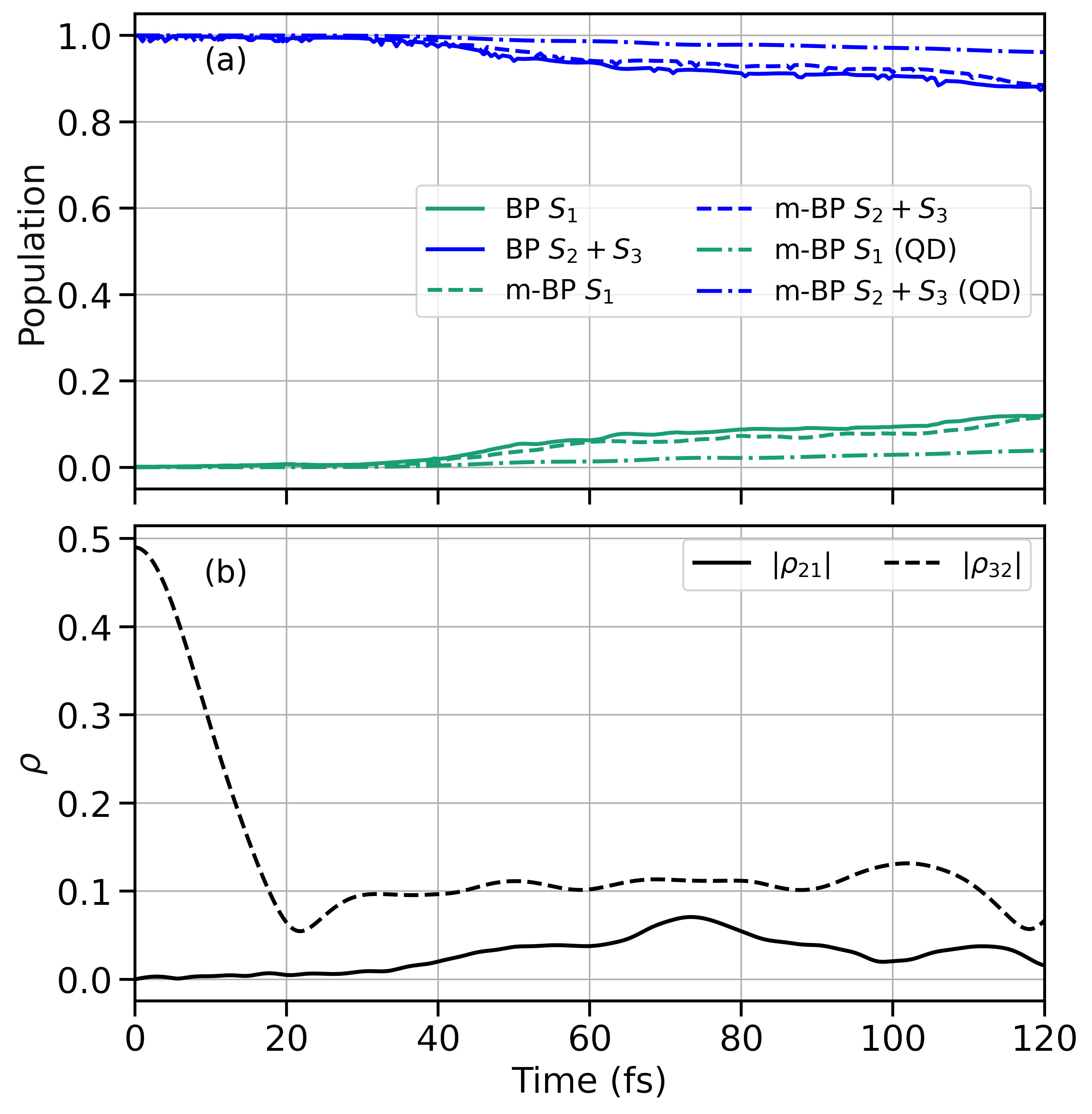}
    \caption{Temporal evolution of diabatic populations and coherences in \gls{bp} and \gls{mbp} during the initial 120~fs. (a): diabatic populations as a function of time, derived from surface hopping (solid and dashed lines) and \gls{qd} (dash-dotted line) simulations. The populations of $S_2$ and $S_3$ are summed together; (b): magnitude of the electronic coherences between $S_2$ and $S_3$ as well as $S_1$ and $S_2$. Subscripts indicate the electronic states involved.}
    \label{QD_SHARC_pops_cohers}
\end{figure}
Figure~\ref{QD_SHARC_pops_cohers} displays the temporal evolution of diabatic populations and electronic coherences, $\rho$, obtained from the \gls{qd} simulations. The chosen reduced reaction coordinate space can capture the general trend for the $S_3$ to $S_2$ internal conversion, as shown in fig.~S10 of the \gls{si}.
However, the $S_3$/$S_2$ dynamics are likely washed out by laser pump-pulses longer than 15~fs.
Moreover, the electronic coherence generated by the $S_3$/$S_2$ \gls{coin} would be largely concealed by the one induced by the pump pulse.
Therefore, the populations of $S_3$ and $S_2$ are summed in fig.~\ref{QD_SHARC_pops_cohers}.
The \gls{qd} simulations can qualitatively reproduce the population transfer from $S_2 + S_3$ to $S_1$ predicted by the surface hopping dynamics.
In particular, the nonadiabatic relaxation into $S_1$ begins at 40~fs and displays a similar linear growth. Despite this, there is a quantitative discrepancy in the total population transfer, as \gls{qd} predicts a 0.05 $S_1$ population at 100~fs,
while the surface hopping dynamics predicts 0.1. A number of factors may
account for this: the underlying electronic structure method used in the two approaches (TDA-LR-DFT vs DFT/MRCI) is different;
the lack of electronic decoherence  in semiclassical trajectory-based dynamics may influence the population transfer directly.
The reaction coordinate space used for \gls{qd} is two-dimensional and does
not include all \gls{dof} that the system may explore during the time interval
of interest.
Beyond 120~fs, these differences become more pronounced as the $S_1$ state is unable to relax within the chosen 2D reaction coordinate space.

\begin{figure}
    \centering
    \includegraphics[width=\linewidth]{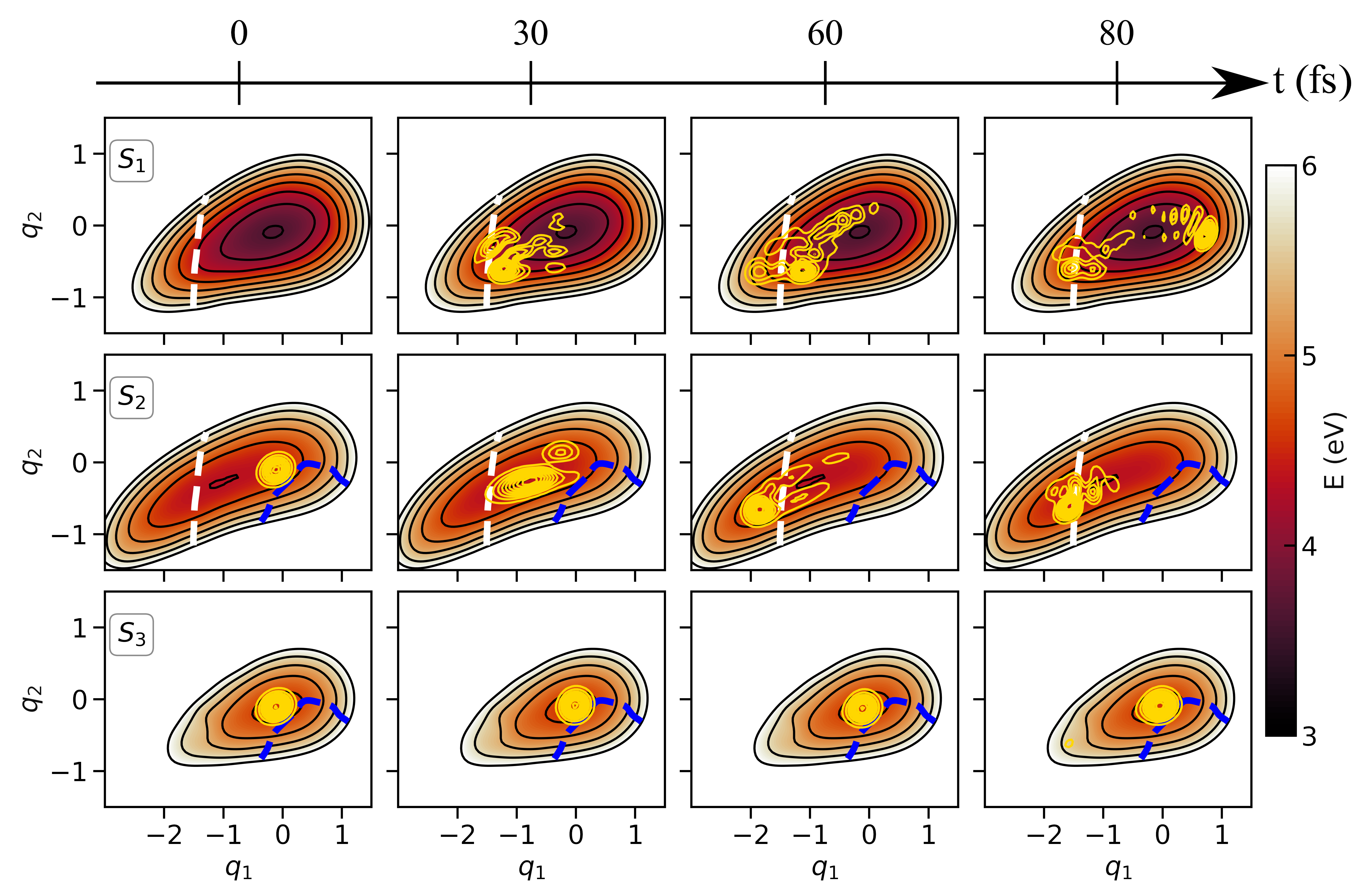}
    \caption{Snapshots of wave packets (gold contour lines) at 0, 30, 60 and 80~fs. First row: wave packet evolution on $S_1$; second row: wave packet evolution on $S_2$; third row: wave packet evolution on $S_3$. The white dashed line indicates the diabatic intersection seam between $S_2$ and $S_1$ including the optimized \gls{coin}. A blue dashed line indicates the diabatic intersection seam between $S_3$ and $S_2$. }
    \label{wave_packet_snap}
\end{figure}
Figure~\ref{wave_packet_snap} shows the time-dependent evolution of the nuclear wave packet on different \glspl{pes}. Snapshots of the nuclear wave packet density ($\vert \Psi \vert ^2$) are captured at $t=0$, 30, 60 and 80~fs.
Since any experimental pump pulse would populate both states, the \gls{qd} simulations start in a coherent superposition of $S_2$ and $S_3$, thus generating a strong electronic coherence $\rho _{32}$ at 0~fs. This coherence rapidly decays over 20~fs, after which it remains relatively constant around 0.1, with minor oscillations, as illustrated in fig.~\ref{QD_SHARC_pops_cohers} (b).
More interesting is the case of $\rho _{21}$ between $S_2$ and $S_1$. In fact, $\rho _{21}$ is zero for the first 35~fs of the dynamics and begins to increase as the wave packet approaches the \gls{coin}, peaking around 70~fs. The magnitude of $\rho_{21}$ is remarkably large and long-lived ($>50$~fs),
especially considering that electronic coherences generated by \glspl{coin} are typically on the order of $10 ^{-3}$ or less~\cite{Neville2022-og} and decay very fast.
Therefore, the calculated 2D \glspl{pes} show promise for following studies where $\rho _{21}$ could be probed using time-resolved X-ray techniques~\cite{Schnappinger2022-pq}, such as TRUECARS~\cite{Kowalewski2015-kv, Restaino2022-jw} or time-resolved X-ray absorption spectroscopy.
Detection of $\rho _{21}$ could serve as a direct spectroscopic signature of the \acrlong{coin}.
\section{Conclusion and Outlook}
The excited-state dynamics of gas-phase benzophenone and its alkyl derivative, \textit{meta}-methyl benzophenone have been investigated using mixed quantum classical and full quantum dynamics simulations, focusing on the nonadiabatic processes in the singlet manifold.
The mixed quantum classical simulations were
carried out at the \acrlong{tda} \acrlong{lrdft} level of theory using a surface hopping approach.
The electronic structure method was benchmarked against
experimental and theoretical data, assuring a qualitatively correct description of all states involved.
We were able to analyze the excited state dynamics for \gls{bp} and \gls{mbp} for 500~fs, observing both the non-adiabatic relaxation process from $S_3$ to $S_2$ and then from $S_2$ to $S_1$.
In addition to discussing the geometrical relaxation and the population dynamics, we also simulate an experimentally accessible observable, namely time-resolved photoelectron spectra for both molecules.

In both cases, we observe an exponential decay of the population in the $S_3$. This behavior is observed in the time-resolved photoelectron spectra as a blueshift of 0.4~eV within the first 15~fs, immediately followed by an interchange in the intensity profile of the photoelectron bands. The population transfer to $S_1$ starts around 40~fs, and the increase/decrease in the population of $S_1$/$S_2$ is linear. As the system passes through the $S_2$/$S_1$ conical intersection, a bifurcation in the photoelectron signal appears in the spectra, followed by a linear increase in the $S_1$ band and a decrease in the intensity of the $S_2$ signal.

To gain more insight and to correctly describe electronic coherences and superpositions of electronic states, we computed two-dimensional potential energy surfaces for \textit{meta}-methylbenzophenone to perform full quantum dynamics simulations.
The presence of the methyl group in the \textit{meta} position increases the energy gap between the $S_3$ to $S_2$ states, which, however, only marginally affects the excited-state dynamics compared to \gls{bp}. This separation is in turn beneficial, as it leads to a slight decoupling of the two states at the Frank-Condon point, allowing for reasonable diabatization.
The quantum dynamics simulations, carried out in a subspace of the full dimensional problem and in a diabatic basis, qualitatively reproduce the surface hopping dynamics over the first 120~fs of propagation.
While the relaxation timescale and linear population increase are accurately depicted, surface hopping and quantum dynamics simulations differ quantitatively.
The former predicts a population of 0.1 in $S_1$ at 100~fs, whereas the latter predicts 0.05. The discrepancy between the two approaches may be attributed to a number of factors, including the differing electronic structure theories employed, and the reduced reaction coordinate space utilized in quantum dynamics.
The electronic coherence generated by the bifurcation of the nuclear wave packet at the $S_2$/$S_1$ conical intersection is remarkably large and survives for more than 50~fs. Consequently, the selected coordinate subspace appears promising for future investigations in which the electronic coherence might be probed with more advanced time-resolved X-ray techniques, which rely on
electronic coherences.

\begin{acknowledgments}
This project has received funding from the European Research Council (ERC) under the European Union’s Horizon 2020 research and innovation program (grant agreement no. 852286).
Support from the Swedish Research Council (Grant No. VR 2022-05005) is acknowledged. L.R. thanks Sambit Das for the many scientific discussions on surface hopping.
\end{acknowledgments}

\section*{Supporting Information}
See the Supporting information for additional benchmarking of vertical excitations of benzophenone using CASSCF and RASSCF; natural orbital plots of the active spaces used in the calculations; normal mode analysis of SHARC trajectories for the first 150~fs; average temporal evolution of selected geometric parameters of benzophenone and \textit{meta}-methyl benzophenone.

\section*{Author Declaration Section}
\subsection*{Conflict of Interest Statement }
The authors have no conflicts to disclose.

\subsection*{Author Contributions}
\textbf{Lorenzo Restaino}: Data curation (lead); Investigation (lead);  Visualization (lead); Writing – original draft (lead); Writing – review \& editing (equal). \textbf{Thomas Schnappinger}: Conceptualization (equal); Investigation (supporting); Supervision (equal);  Writing – original draft (supporting); Writing – review \& editing (equal). \textbf{Markus Kowalewski}: Conceptualization (equal); Funding
acquisition (lead);
Project administration (lead); Supervision (equal); Writing – original draft (supporting); Writing – review \& editing (equal).

\section*{Data Availability}
The data that support the findings of this study are available from the corresponding author upon reasonable request.

\bibliography{Benzophenone.bib}

\bibliographystyle{unsrt}

\end{document}

\title{Supporting Informations: Simulating Nonadiabatic Dynamics in Benzophenone: Tracing Internal Conversion Through Photoelectron Spectra}

\author{Lorenzo Restaino}
\affiliation{Department of Physics, Stockholm University, Albanova University Centre, SE-106 91 Stockholm, Sweden}
\author{Thomas Schnappinger}%
\affiliation{Department of Physics, Stockholm University, Albanova University Centre, SE-106 91 Stockholm, Sweden}
\author{Markus Kowalewski}%
\email{e-mail: markus.kowalewski@fysik.su.se}
\affiliation{Department of Physics, Stockholm University, Albanova University Centre, SE-106 91 Stockholm, Sweden}

\maketitle
\tableofcontents
\clearpage

\section{Additional Benchmark of Vertical Excitation Energies}
\subsection{CASPT2}
Benzophenone in its ground state is a non-planar molecule with $C_2$ symmetry. The steric repulsion of hydrogen atoms causes rotation between the phenyl rings. We noticed that, starting from the actual twisted structure, the calculations produce incorrect $S_0$ and $S_1$ state energies. To correct for this, we first carried out a single-point CASSCF calculation on the planar structure and then used the resulting RasOrb file on the twisted ground-state structure.
\begin{table}[h!]
\caption{Additional benchmark of vertical energies (eV) for gas-phase benzophenone.}
\begin{tabular}{cccccccccc}
\hline
\multicolumn{1}{l}{} & \multicolumn{2}{c}{\textbf{CASPT2(12,11)}} & \multicolumn{2}{c}{\textbf{CASPT2(14,13)}} & \multicolumn{2}{c}{\textbf{CASPT2(16,15)}} & \multicolumn{1}{c}{\textbf{CCSD}$^a$} & \multicolumn{1}{c}{\textbf{TDA}$^b$} & \multicolumn{1}{c}{\textbf{Exp.}} \\ \hline
\multicolumn{1}{l}{} & \textbf{4avg} & \textbf{5avg} & \textbf{4avg} & \textbf{5avg} & \textbf{4avg} & \textbf{5avg} & \multicolumn{1}{c}{\textbf{}} & \multicolumn{1}{l}{} & \multicolumn{1}{c}{} \\ \hline
$S_0 \rightarrow S_1$ & 3.57 & 3.66 & 3.57 & 3.65 & 3.53 & 3.65 & 3.69 & 3.98 & 3.61$^c$ \\ 
$S_0 \rightarrow S_2$ & 4.96 & 5.20 & 4.59 & 4.59 & 4.60 & 4.60 & 4.60 & 5.28 & 4.40$^c$ \\ 
$S_0 \rightarrow S_3$ & 4.99 & 5.24 & 4.89 & 4.88 & 4.67 & 4.62 & 4.67 & 5.32 & 4.40 \\ 
$S_0 \rightarrow S_4$ &  & 5.56 &  & 5.83 &  & 5.83 & 5.47 & 5.62 & 5.00 \\ \hline
\multicolumn{10}{l}{$^a$ STEOM-DLPNO-CCSD/def2-TZVP; $^b$ $\omega$B97-D3/6-311G*; $^{c}$ low pressure vapor, ref.~\cite{Itoh1985-jk}.}
\end{tabular}
\label{Table1}
\end{table}
\begin{table}[h!]
 \caption{Oscillator strength of the vertical transitions reported in table~\ref{Table1}.}
\begin{tabular}{cccccccccc}
\hline
\multicolumn{1}{l}{} & \multicolumn{2}{c}{\textbf{CASPT2(12,11)}} & \multicolumn{2}{c}{\textbf{CASPT2(14,13)}} & \multicolumn{2}{c}{\textbf{CASPT2(16,15)}} & \multicolumn{1}{c}{\textbf{CCSD}$^a$} & \multicolumn{1}{c}{\textbf{TDA}$^b$} & \multicolumn{1}{c}{\textbf{Exp.}} \\ \hline
\multicolumn{1}{l}{} & \textbf{4avg} & \textbf{5avg} & \textbf{4avg} & \textbf{5avg} & \textbf{4avg} & \textbf{5avg} & \multicolumn{1}{c}{\textbf{}} & \multicolumn{1}{l}{} & \multicolumn{1}{c}{} \\ \hline
$S_0 \rightarrow S_1$ & 0.0009 & 0.0010 & 0.0010 & 0.0009 & 0.0009 & 0.0007 & 0.0008 & 0.0011 & \textbf{weak} \\ 
$S_0 \rightarrow S_2$ & 0.0048 & 0.0062 & 0.0021 & 0.0033 & 0.0014 & 0.0036 & 0.0080 & 0.0165 & \multicolumn{1}{l}{} \\ 
$S_0 \rightarrow S_3$ & 0.0034 & 0.0104 & 0.0044 & 0.0059 & 0.0027 & 0.0023 & 0.0058 & 0.0164 & \multicolumn{1}{l}{} \\ 
$S_0 \rightarrow S_4$ &  & 0.0032 &  & 0.0145 &  & 0.0068 & 0.5067 & 0.4387 & \textbf{strong} \\ \hline
\multicolumn{10}{l}{$^a$ $\omega$B97-D3/6-311G*; $^b$ STEOM-DLPNO-CCSD/def2-TZVP.}
\end{tabular}
\label{Table2}
\end{table}
\begin{figure}
    \centering
\includegraphics[width=0.7\linewidth]{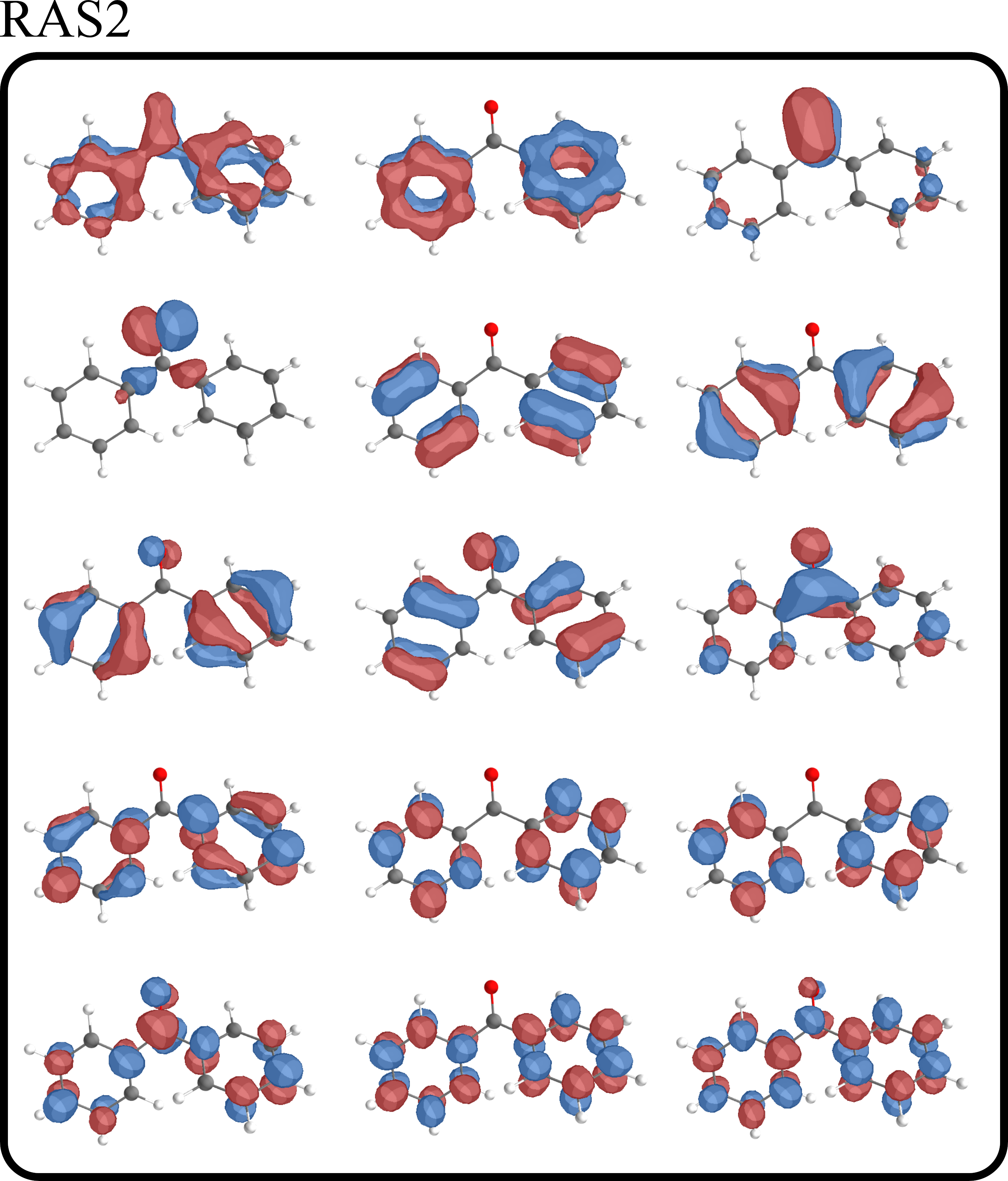}
    \caption{Natural orbitals of benzophenone in the active space with 16 electrons in 15 orbitals, calculated at the CAS(16,16)/ANO-L-VDZP level of theory.}
    \label{active_space16_15}
\end{figure}

Sergentu et al.~\cite{Sergentu2014-da} calculated a minimum energy path (MEP) at the CASPT2 level on a CAS(12,11) wave function with the ANO-L-VDZP basis set and stated that it does not lead to the $S_2$/$S_1$ conical intersection reporting an energy splitting of 0.69 eV. They concluded that CAS(12,11) is inadequate to describe the state $S_2$ due to the absence of the lowest $\pi$ orbitals and corresponding $\pi ^*$ in the active space. We agree that an accurate characterization of the $S_2$ PES is crucial for understanding the photoexcitation dynamics, though our results differ slightly from their findings. In their paper, they did not specify which CASPT2 variant was used, but our replication of their results was successful only with multi-state (MS) CASPT2. 

We employed extended multi-state CASPT2 (XMS-CASPT2), which provides a better description of the states, particularly when state mixing is strong. Our benchmarks for gas-phase benzophenone vertical energies are in table~\ref{Table1}.
The analysis of the CI coefficients for $S_2$, $S_3$, and $S_4$ did not show significant contributions from the two lowest $\pi$ orbitals, suggesting that the improved description of $S_2$ is simply due to a larger active space, where the correlation of the $\pi$-system is better. This is evident in tables~\ref{Table1} and~\ref{Table3}. The $S_1$ state is unaffected by the active space size, but slightly influenced by the state-averaging number, with differences under 0.1 eV. The smaller active space (12,11) inadequately describes $S_2$, showing differences over 0.4 eV compared to the largest active space. So the conclusion reached by Sergentu et al. regarding this state still holds true.

\clearpage
\newpage
\subsection{RASPT2}
We have also benchmarked vertical excitations in the single manifold with RASPT2~\cite{Sauri2011-au,Malmqvist2008-ol}. The energies and the corresponding oscillator strength are reported in tables~\ref{Table3}. and~\ref{Table4}. 
\begin{table}[h!]
\caption{Benchmark of vertical energies (eV) for gas-phase benzophenone using XMS-RASPT2, with 2 orbitals in the RAS1, 11 orbitals in the RAS2 and 2 orbitals in the RAS3. One hole and one particle are allowed in the RAS1 and RAS3, respectively.}
\begin{tabular}{cccccc}
\hline
 & \multicolumn{2}{c}{\textbf{RASPT2}} &  \multicolumn{2}{c}{\textbf{CASPT2(16,15)}} & \multicolumn{1}{c}{\textbf{Exp.}} \\ \hline
& \textbf{4avg} & \textbf{5avg}  & \textbf{4avg} & \textbf{5avg} &  \\ \hline
$S_0 \rightarrow S_1$ & 3.83 & 3.65  & 3.53 & 3.65 & 3.61 \\ 
$S_0 \rightarrow S_2$ & 4.61 & 4.67 & 4.60 & 4.60 & 4.40 \\ 
$S_0 \rightarrow S_3$& 4.63 & 4.67  & 4.67 & 4.62 & 4.40 \\ 
$S_0 \rightarrow S_4$ & & 5.80  &  & 5.83 & 5.00 \\ \hline
\end{tabular}
\label{Table3}
\end{table}
\begin{table}[h!]
 \caption{Oscillator strength of the vertical transitions reported in table~\ref{Table3}.}
\begin{tabular}{cccccc}
\hline
& \multicolumn{2}{c}{\textbf{RASPT2}} & \multicolumn{2}{c}{\textbf{CASPT2(16,15)}} & \multicolumn{1}{c}{\textbf{Exp.}} \\ \hline
& \textbf{4avg} & \textbf{5avg}  & \textbf{4avg} & \textbf{5avg} &  \\ \hline
$S_0 \rightarrow S_1$ & 0.0013
& 0.0009   & 0.0009 & 0.0007 & \textbf{weak} \\ 
$S_0 \rightarrow S_2$ & 0.0042
& 0.0049  & 0.0014 & 0.0036 &  \\ 
$S_0 \rightarrow S_3$ & 0.0017
 & 0.0032 &  0.0027 & 0.0023 &  \\ 
$S_0 \rightarrow S_4$ & & 0.0082 &  & 0.0068 & \textbf{strong} \\ \hline
\end{tabular}
\label{Table4}
\end{table}
 \begin{figure}[hbt]
    \centering
    \includegraphics[width=0.7\linewidth]{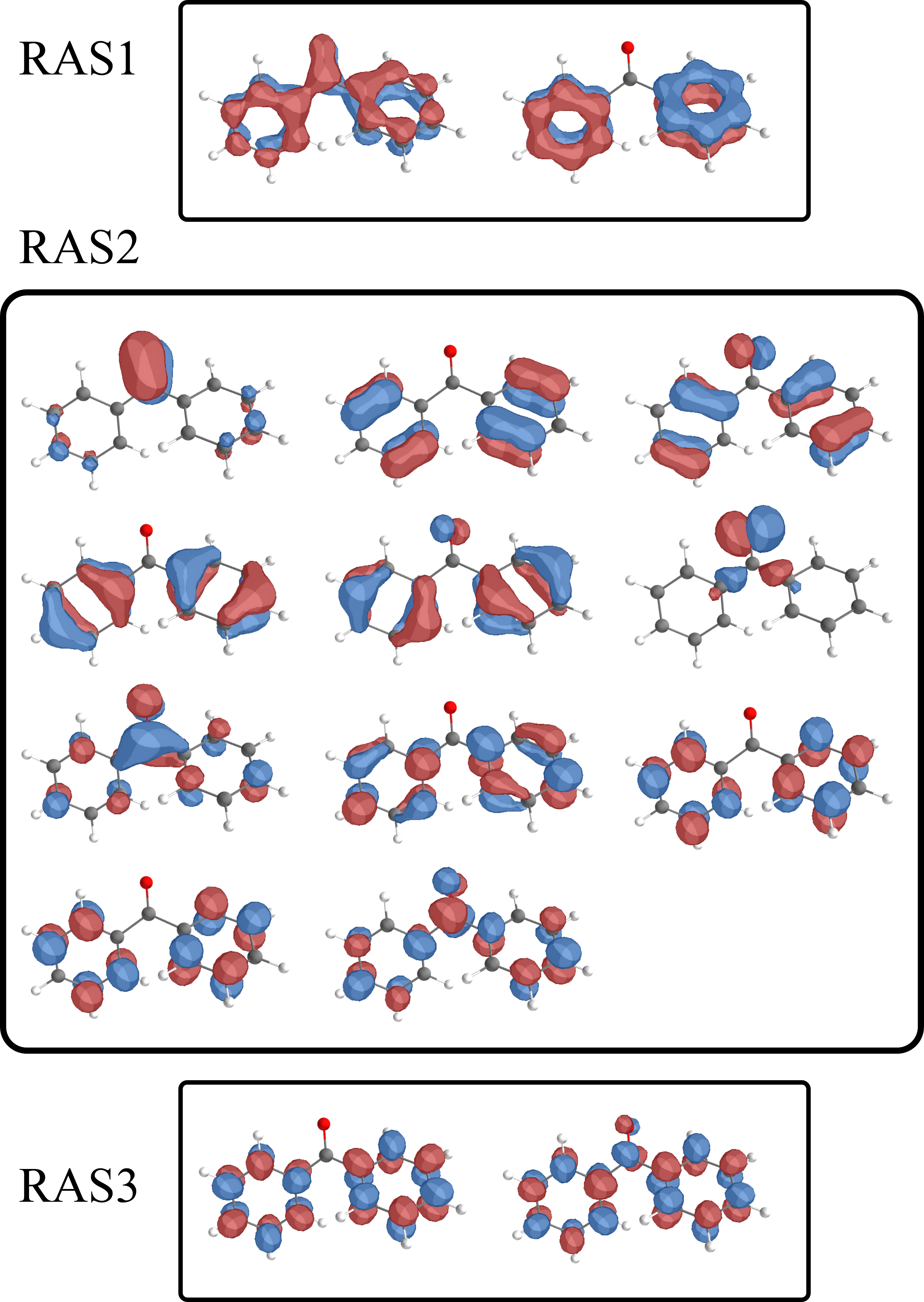}
    \caption{Natural orbitals of benzophenone in the restricted active space formalism with two orbitals in the RAS1, 11 orbitals in the RAS2, and two orbitals in the RAS3. 12 electrons in the RAS2, one hole and one particle allowed in the RAS1 and RAS3 respectively.}
    \label{active_space_ras}
\end{figure}

\clearpage
\newpage
\section{DFT/MRCI Active Space}
\begin{figure}[hbt]
    \centering
\includegraphics[width=0.7\linewidth]{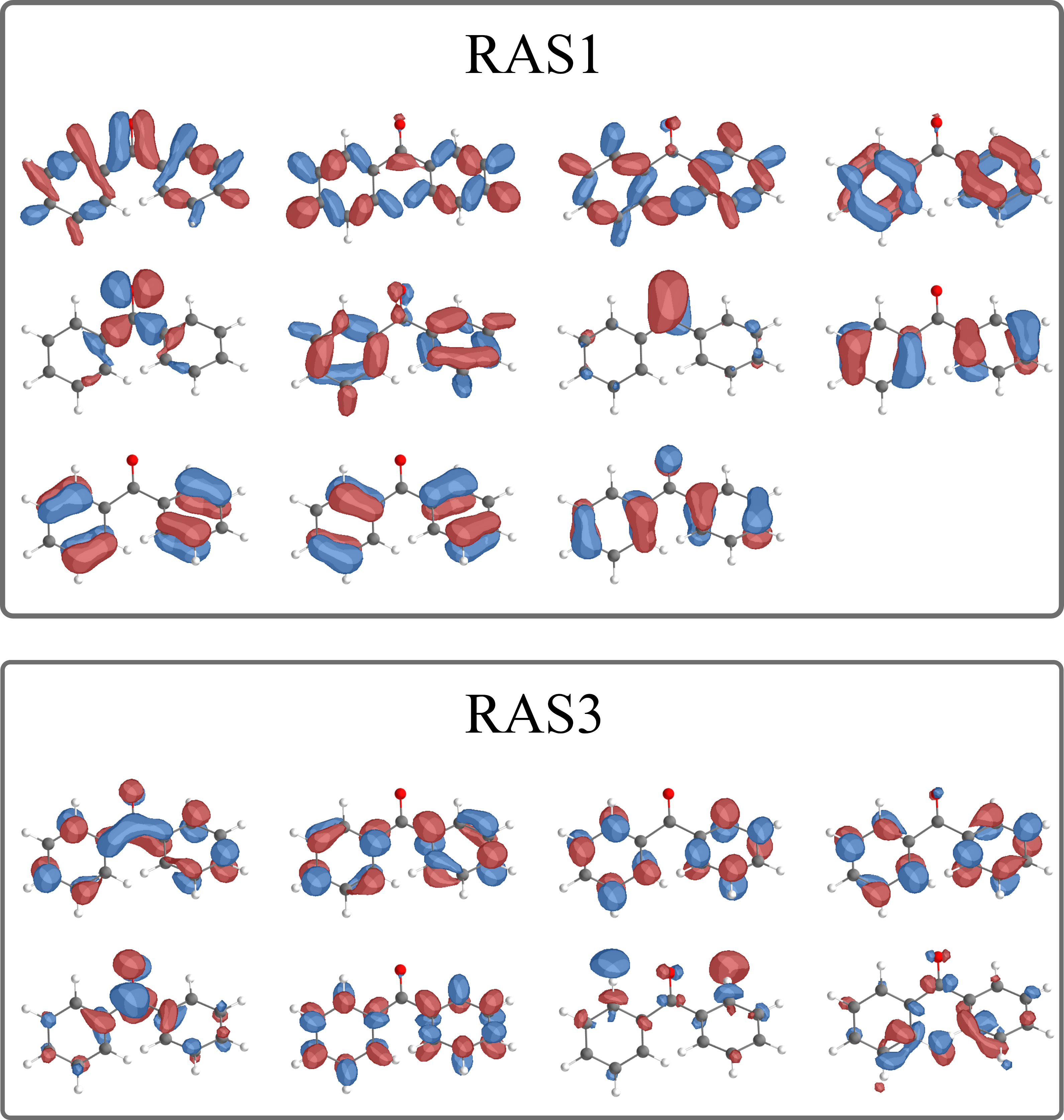}
    \caption{Natural orbitals of benzophenone in the active space used in DFT/MRCI calculations. Two holes and two particles are allowed in the RAS1 and RAS3 respectively.}
    \label{DFT_MRCI_active_space_BP}
\end{figure}

\begin{figure}[hbt]
    \centering
\includegraphics[width=0.7\linewidth]{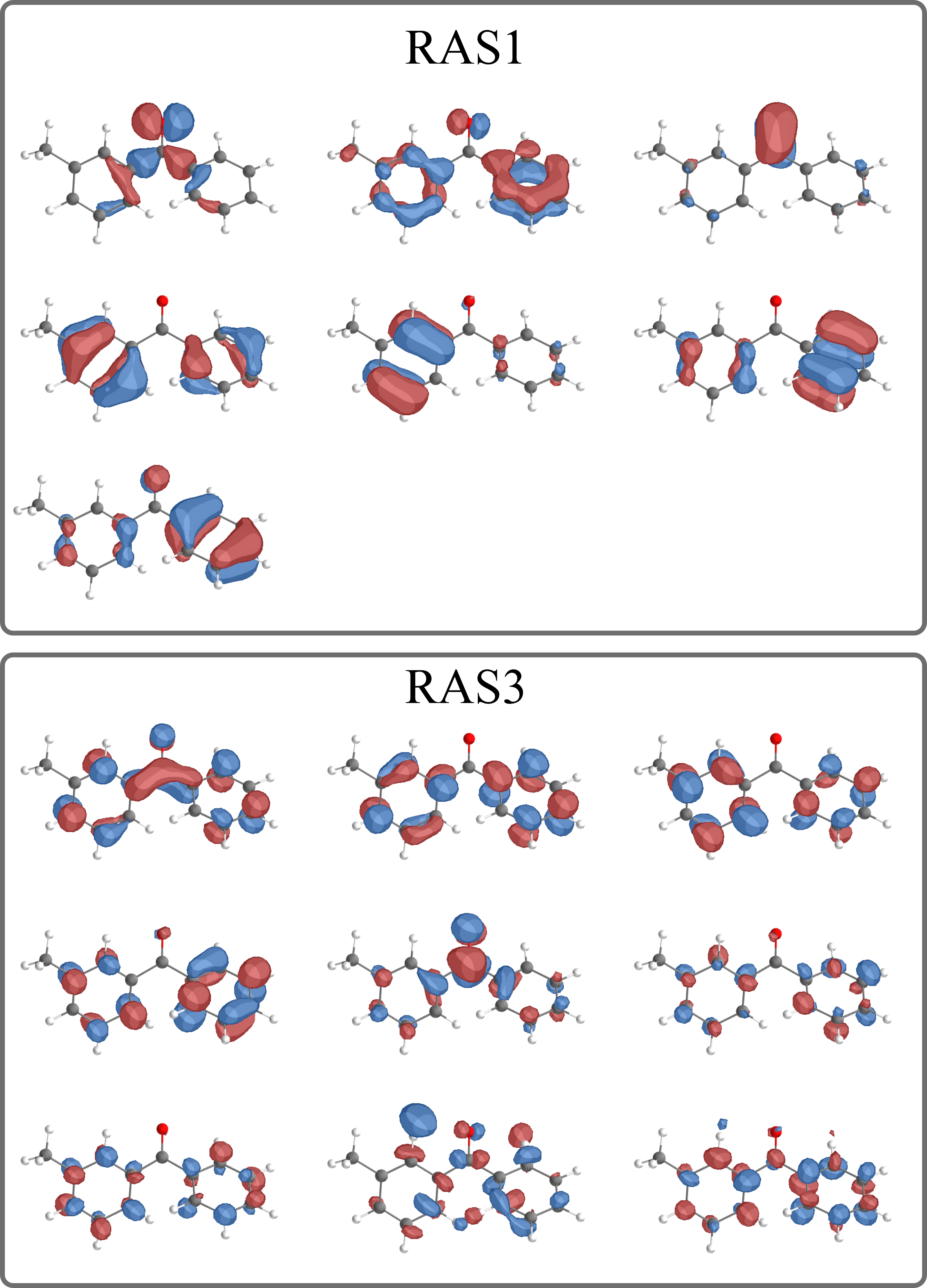}
    \caption{Natural orbitals of \textit{meta}-methyl benzophenone in the active space used in DFT/MRCI calculations. Two holes and two particles are allowed in the RAS1 and RAS3 respectively.}
    \label{DFT_MRCI_active_space_mBP}
\end{figure}
\clearpage
\newpage

\clearpage
\newpage
\section{Normal Mode Analysis with SHARC}
The activity of the normal modes of BP and m-BP for their respective ensemble of trajectories was calculated as implemented in SHARC~\cite{Kurtz2001-lg}. We restrict our analysis to the first 150 fs of the dynamics. 
\begin{figure}[hbt]
    \centering
    \includegraphics[width=0.6\linewidth]{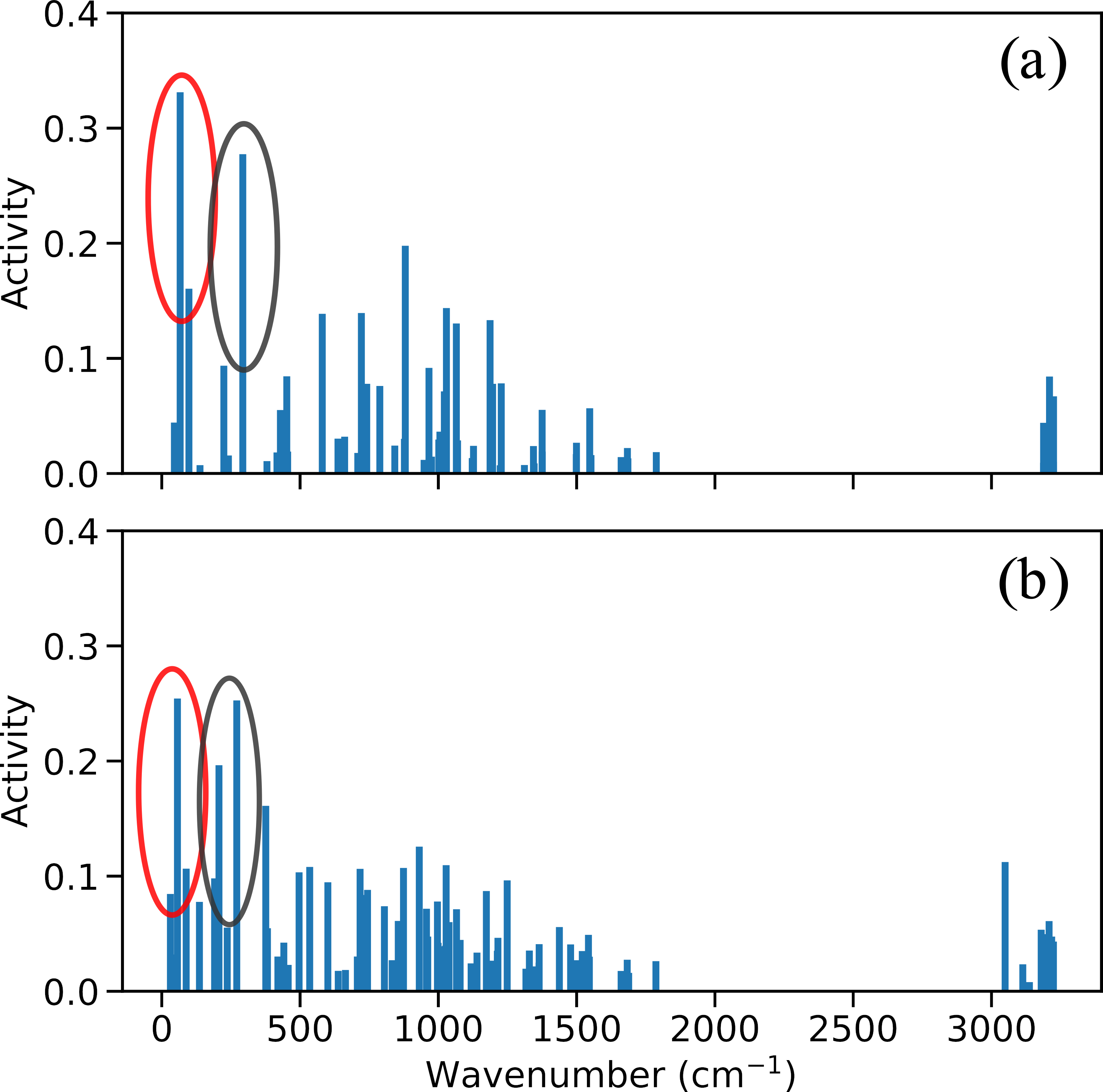}
    \caption{Normal mode analysis of the average trajectory for (a) benzophenone and (b) \textit{meta}-methyl benzophenone, for the first 150 fs of the dynamics. Red oval indicate normal modes associated with the dihedral motion of the two aryl rings. Grey oval indicate normal modes associated with C-C stretching involving the carbon atom of the carbonyl group.}
    \label{NMA_analysis}
\end{figure}
\begin{figure}[hbt]
    \centering
    \includegraphics[width=\linewidth]{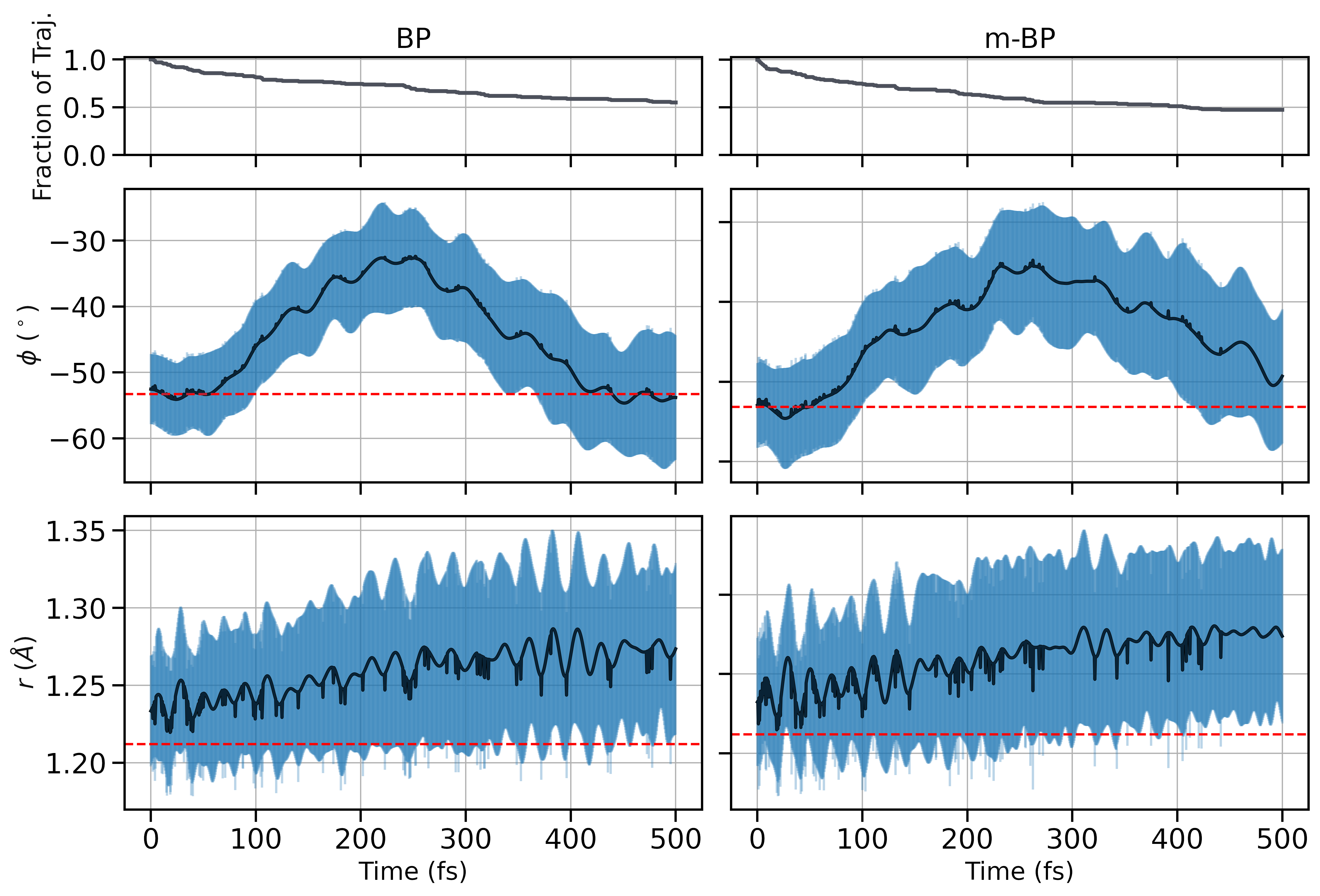}
    \caption{Mean progression of geometric parameters over time for BP (left column) and m-BP (right column). Top panel: fraction of trajectories (grey line) meeting the total energy conservation criteria at each time step; middle panel: mean temporal progression of the dihedral angle between the two aryl rings; bottom panel: mean temporal progression of the CO bond length. Red dashed lines indicate values at the FC point.}
    \label{}
\end{figure}

\clearpage
\newpage

\section{Potential Energy surfaces}
\begin{figure}[hbt]
    \centering
\includegraphics[width=0.9\linewidth]{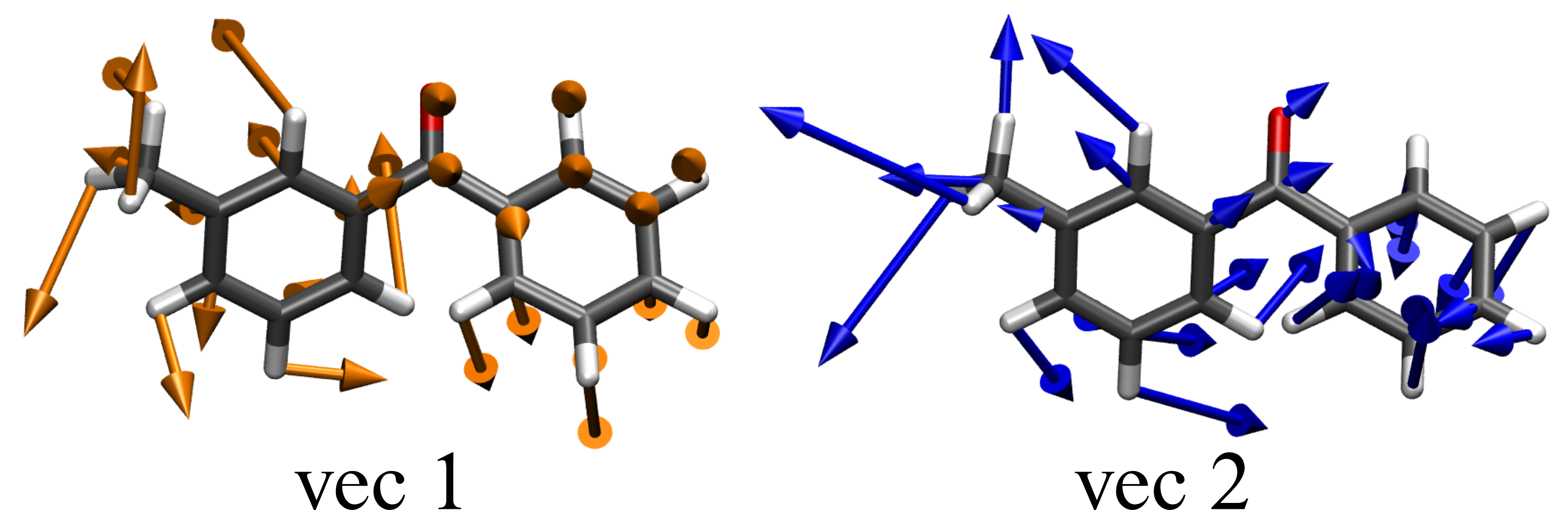}
    \caption{Graphical depiction of the orthonormalized vectors spanning the coordinate space used to form potential energy surfaces for m-BP. Vec1 represents the difference vector between the FC geometry and the CoIn geometry. Vec2 represents the dihedral angle between the aryl rings. }
    \label{meta_BZP_vecs_PES}
\end{figure}

\begin{figure}[hbt]
    \centering
\includegraphics[width=0.8\linewidth]{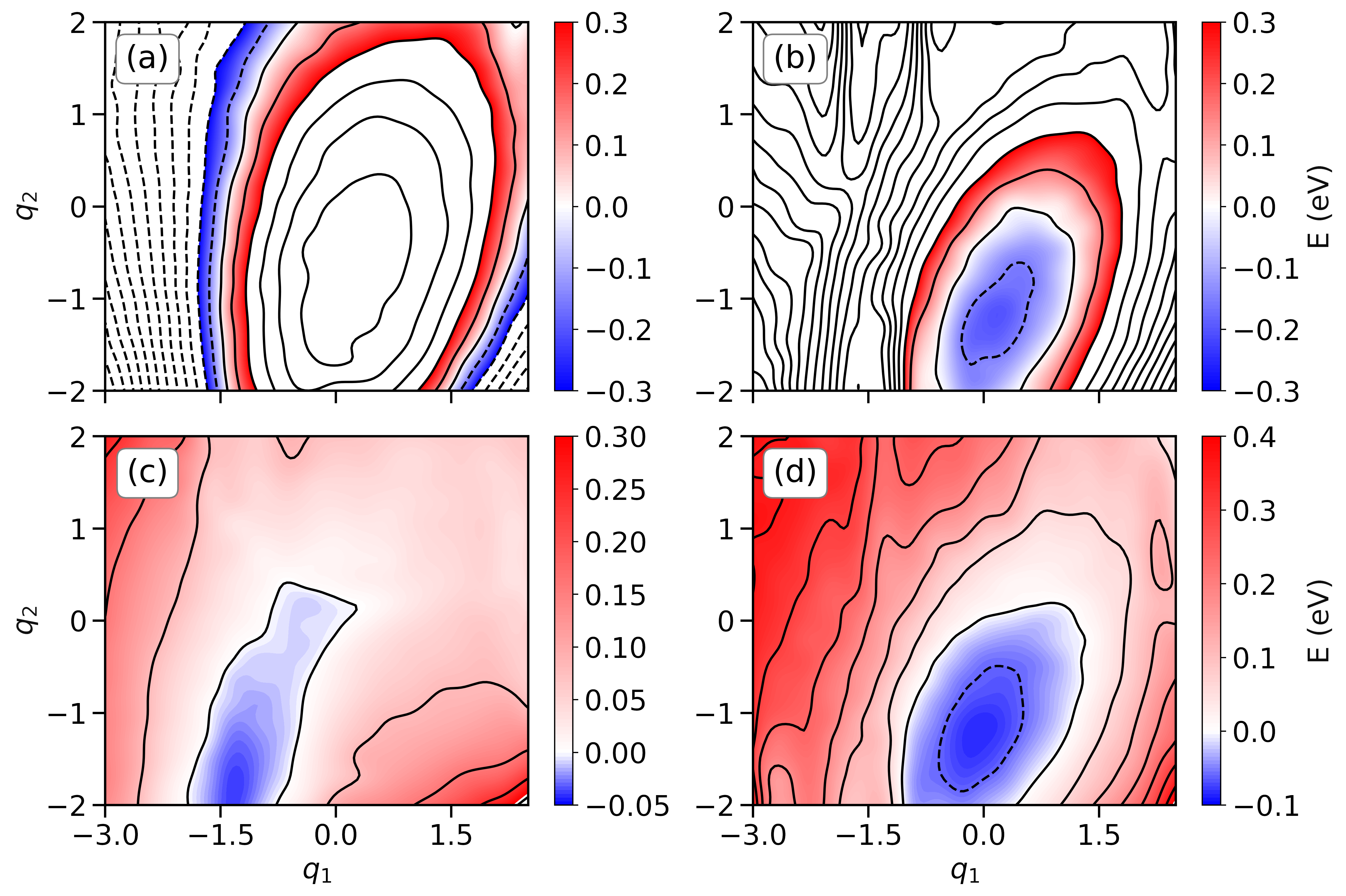}
    \caption{Diabatic potential energy differences and diababtic couplings: (a) $S_2 - S_1$; (b) $S_3 - S_2$; (c) $S_2$/$S_1$ diabatic coupling; (d) $S_3$/$S_2$ diabatic coupling.}
    \label{meta_BP_PES_diff}
\end{figure}
\clearpage
\newpage
\section{SHARC Populations of BP and m-BP}
\begin{figure}[hbt]
    \centering
    \includegraphics[width=0.6\linewidth]{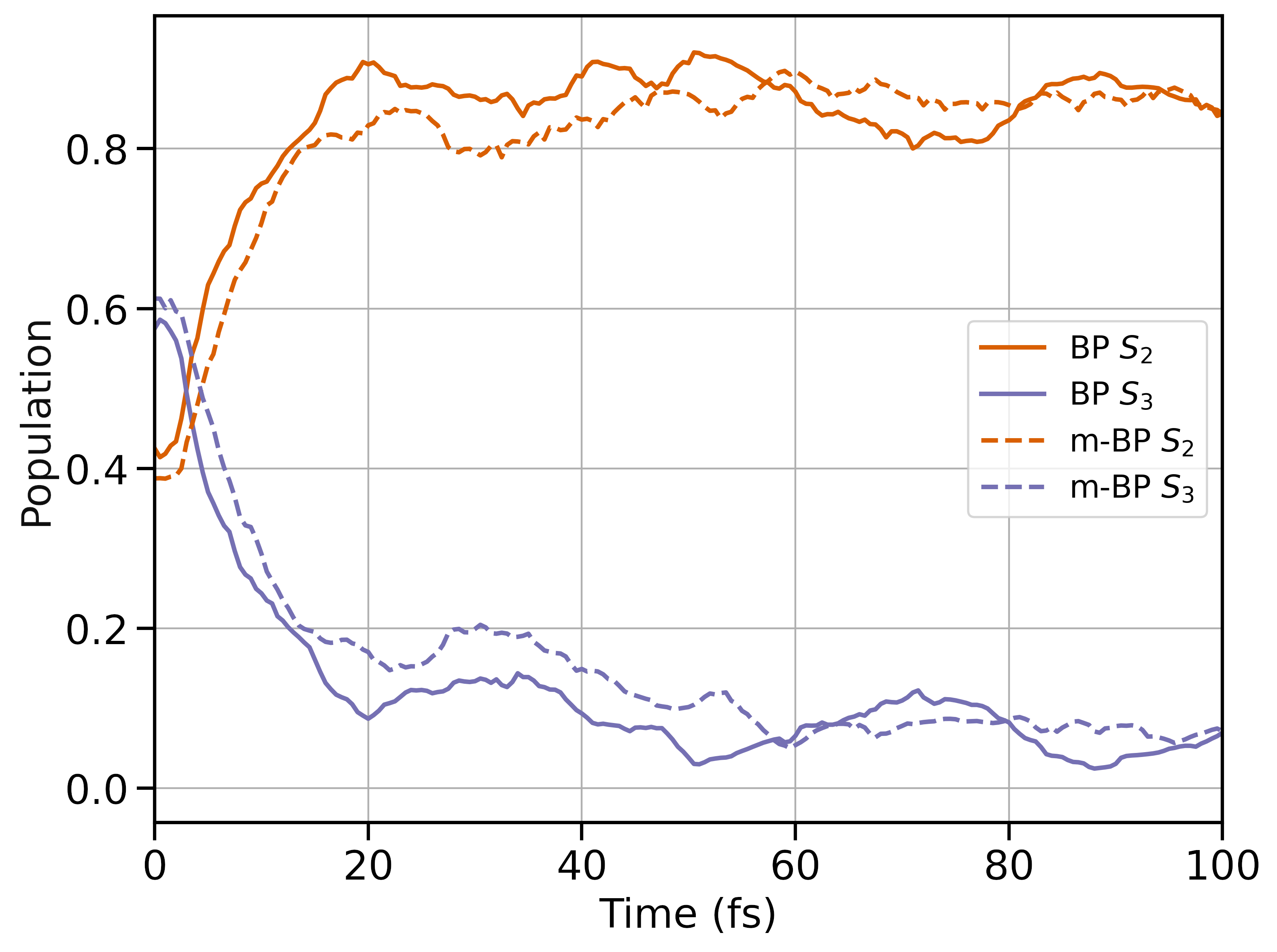}
    \caption{Comparison of the adiabatic SHARC populations of the $S_2$ and $S_3$ states of benzophenone (solid lines) and \textit{meta}-methyl benzophenone (dashed lines), within the initial 100~fs. }
    \label{sharc_pop_100fs}
\end{figure}
\begin{figure}
    \centering
\includegraphics[width=0.65\linewidth]{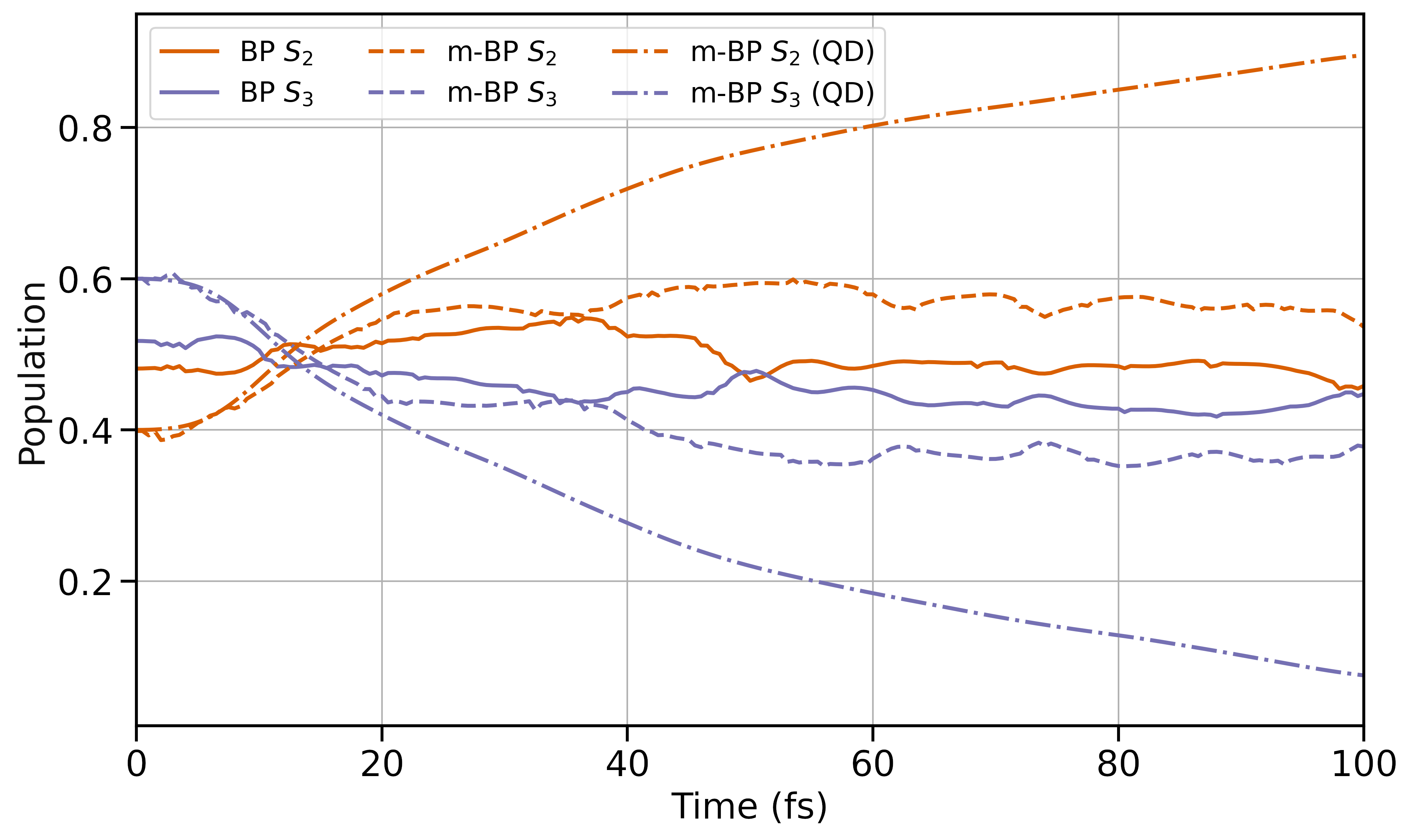}
    \caption{Temporal evolution of diabatic SHARC populations for the $S_2$ and $S_3$ states in benzophenone (solid lines) and \textit{meta}-methyl benzophenone (dashed lines), alongside diabatic populations of \textit{meta}-methyl benzophenone derived from QD simulations (dashed-dotted lines).}
    \label{fig:enter-label}
\end{figure}
\clearpage
\newpage
\bibliography{Benzophenone.bib}

\bibliographystyle{unsrt}